\documentclass[pdflatex,sn-mathphys-num]{sn-jnl}% Math and Physical Sciences Numbered Reference Style 
\usepackage{graphicx}%
\usepackage{multirow}%
\usepackage{multicol}
\usepackage{amsmath,amssymb,amsfonts}%
\usepackage{amsthm}%
\usepackage{mathrsfs}%
\usepackage[title]{appendix}%
\usepackage{xcolor}%
\usepackage{textcomp}%
\usepackage{manyfoot}%
\usepackage{booktabs}%
\usepackage{algorithm}%
\usepackage{algorithmicx}%
\usepackage{algpseudocode}%
\usepackage{listings}%
\usepackage{subcaption}
\usepackage{graphicx}
\usepackage{amsmath}
\usepackage{array}
\usepackage{amssymb}
\usepackage{amsmath}
\usepackage{multirow}
\usepackage{subcaption}
\usepackage{xcolor}
\usepackage{rotating}
\usepackage{wrapfig}
\usepackage{float}  % Add this to your preamble
\usepackage{longtable}

\theoremstyle{thmstyleone}%
\theoremstyle{thmstyletwo}%

\theoremstyle{thmstylethree}%

\begin{document}

\title[How good nnU-Net for Segmenting Cardiac MRI: A Comprehensive
Evaluation]{How good nnU-Net for Segmenting Cardiac MRI: A Comprehensive Evaluation}


% \author*[1,2]{\fnm{First} \sur{Author}}\email{iauthor@gmail.com}



% \affil*[1]{\orgdiv{Department}, \orgname{Organization}, \orgaddress{\street{Street}, \city{City}, \postcode{100190}, \state{State}, \country{Country}}}

% \affil[2]{\orgdiv{Department}, \orgname{Organization}, \orgaddress{\street{Street}, \city{City}, \postcode{10587}, \state{State}, \country{Country}}}

% \affil[3]{\orgdiv{Department}, \orgname{Organization}, \orgaddress{\street{Street}, \city{City}, \postcode{610101}, \state{State}, \country{Country}}}

\author*[1]{\fnm{Malitha} \sur{Gunawardhana}}\email{malithagunawardhana96@gmail.com}
\author[1]{\fnm{Fangqiang} \sur{Xu}}
\author[1]{\fnm{Jichao} \sur{Zhao}}

\affil*[1]{\orgdiv{Auckland Bioengineering Institute}, \orgname{University of Auckland}, \orgaddress{ \city{Auckland}, \postcode{1010}, \country{New Zealand}}}

\abstract{  
\textbf{Background}  

Cardiac segmentation is a critical process in medical imaging, providing a detailed analysis of heart structures that is essential for the diagnosis and treatment of various cardiovascular diseases. The advent of deep learning has revolutionized this field by introducing automated segmentation techniques that significantly outperform traditional manual methods in terms of both accuracy and efficiency. The nnU-Net framework, in particular, has emerged as a highly robust and versatile tool for medical image segmentation, capable of adapting to a wide range of imaging modalities and segmentation tasks.

\textbf{Methods}  

In this study, we systematically evaluate the performance of nnU-Net in segmenting cardiac magnetic resonance images (MRIs) using five distinct datasets: LAScarQs 2022, LASC 2018, ACDC, MnM1, and MnM2. We investigate the efficacy of various nnU-Net configurations, including 2D, 3D full resolution, 3D low resolution, 3D cascade, and ensemble models. Each configuration is assessed based on its ability to accurately delineate cardiac structures, with a focus on comparing their strengths and weaknesses across different cardiac segmentation tasks. The study employs a comprehensive benchmarking approach to evaluate the models' capabilities and identify potential areas for improvement.

\textbf{Results}  

Our findings indicate that the nnU-Net configurations deliver state-of-the-art performance across all the datasets examined. Notably, the 2D configuration outperformed the 3D configuration in certain scenarios, demonstrating superior accuracy and efficiency. In some cases, the 2D model even surpassed other advanced approaches. However, ensemble methods did not consistently provide additional benefits, highlighting that more complex models are not always superior. The analysis also reveals specific limitations in existing models, particularly in handling challenging segmentation tasks, underscoring the need for further refinement.

\textbf{Conclusions}  

This study highlights the robustness and flexibility of nnU-Net configurations in cardiac MRI segmentation, affirming their effectiveness. However, the results also emphasize the need for developing new, specialized models to address particular segmentation challenges. These insights are valuable for guiding future research aimed at improving deep learning methodologies for more precise and efficient cardiac imaging analysis.
}


\maketitle

\section{Introduction}

Cardiovascular diseases (CVDs) accounted for an estimated 19.05 million deaths globally in 2020, reflecting an 18.71\% increase from 2010. Despite this rise, the age-standardized death rate decreased by 12.19\%, reaching 239.80 per 100,000 population. Additionally, the total crude prevalence of CVD worldwide reached 607.64 million cases in 2020, marking a 29.01\% increase compared to 2010~\citep{tsao2023heart}. These statistics underscore the urgent need for advanced diagnostic and therapeutic approaches in cardiology. 

Accurate segmentation of cardiac structures is essential for understanding heart function, planning interventions, and monitoring disease progression. For example, locating and quantifying fibrosis and scars have been demonstrated to be valuable tools for the treatment stratification of patients with atrial fibrillation (AF)~\citep{allessie2002electrical,boldt2004fibrosis,gunawardhana2024integrating} and ventricular tachycardia~\citep{ukwatta2015image}. These techniques provide critical guidance for surgical or ablation procedures~\citep{vergara2011tailored}, and imaging of post-ablation scars offers valuable insights into treatment outcomes~\citep{peters2007detection}.

Cardiac segmentation involves the precise delineation of key anatomical structures within the heart, including the myocardium, ventricles, atria, and major vessels. In particular, Late Gadolinium Enhancement Magnetic Resonance Imaging (LGE-MRI) has emerged as an invaluable technique in cardiac imaging. LGE-MRI excels in highlighting areas of myocardial scarring and fibrosis, which are critical indicators in the diagnosis and management of various cardiac conditions, including myocardial infarction, cardiomyopathy, and arrhythmia~\citep{akkaya2013relationship,bisbal2014cmr}.

Historically, manual segmentation by expert radiologists and cardiologists has been considered the gold standard for cardiac image analysis. However, this method is hindered by significant limitations, particularly its time-consuming nature, often requiring hours for a single dataset, making it impractical for busy clinical settings~\citep{tobon2015benchmark}. The advent of automated segmentation methods, especially those utilizing deep learning techniques, has transformed the field of cardiac imaging analysis. These methods offer substantial advantages over traditional manual approaches, including consistency and reproducibility by eliminating inter-observer variability, rapid analysis with deep learning models capable of segmenting cardiac structures within seconds, scalability for application to large datasets, and the potential for continuous improvement as models can be fine-tuned and updated with increasing data availability.

Over the past decade, numerous approaches have been developed for automated cardiac segmentation, each with its own strengths and limitations. These methods have explored various approaches to improve segmentation accuracy and robustness, including utilizing uncertainty~\citep{yang2019combating,arega2022using}, semi-supervised learning~\citep{shi2024mlc,mazher2022automatic}, curriculum learning~\citep{jiang2022deep}, and multi-task learning~\citep{chen2019multi}.

Despite these advancements, there remains a notable gap in the literature regarding the comprehensive evaluation of one particular architecture that has shown remarkable success in medical image segmentation across various domains: the nnU-Net (no-new-Net)~\citep{isensee2021nnu}. The nnU-Net is a self-configuring method based on the U-Net architecture that automatically adapts preprocessing, network architecture, training, and post-processing to the specifics of a given dataset. While nnU-Net has demonstrated state-of-the-art performance in numerous biomedical segmentation challenges~\citep{isensee2024nnu}, its potential in the specific context of cardiac segmentation has not been thoroughly explored. This presents a significant research opportunity, as cardiac MRI poses unique challenges due to its high contrast between normal and scarred myocardium, potential artefacts, and variability in image quality across different scanners and institutions.

In this study, we aim to bridge this knowledge gap by conducting a comprehensive analysis of nnU-Net's performance in segmenting cardiac MRI. We utilize five widely used datasets for this task. Those are LAScarQS 2022 dataset ~\citep{zhuang2023left}, 2018 LASC dataset ~\citep{xiong2021global},
ACDC dataset~\citep{bernard2018deep}, MnM ~\citep{campello2021multi} and MnM2 datasets~\citep{martin2023deep}. To the best of our knowledge, this is the first study to focus exclusively on this combination of methodology and imaging modality. By conducting this comprehensive analysis, we aim to provide the medical imaging community with valuable insights into the capabilities and limitations of nnU-Net for segmenting cardiac MRI. Our findings could potentially influence future directions in algorithm development, clinical adoption of automated segmentation tools, and standardization efforts in cardiac imaging analysis.

% The remainder of this paper is organized as follows: Section 2 provides a detailed background on the nnU-Net architecture. Section 3 describes our methodology, including dataset preparation, experimental setup, and evaluation metrics. Section 4 presents our results and analysis. Section 5 discusses the implications of our findings, the limitations of the study, and future research directions. Finally, Section 6 concludes the paper by summarizing the key contributions and insights gained from this research, including an answer to the question of when it is necessary to develop new models specifically tailored for particular cardiac segmentation tasks.

\section{Method}

\subsection{nnU-Net architecture}

The nnU-Net framework is specifically designed for semantic segmentation and is capable of handling both 2D and 3D images with various input modalities or channels. It adeptly processes voxel spacing and anisotropies and exhibits robustness even in scenarios where class distributions are highly imbalanced. Utilizing supervised learning, nnU-Net necessitates the provision of annotated training cases tailored to the application at hand. The quantity of required training cases can vary significantly depending on the complexity of the segmentation task, though nnU-Net often requires fewer cases than other solutions due to its extensive data augmentation strategies.

A key expectation for nnU-Net is its ability to process entire images during both the preprocessing and post-processing stages, making it unsuitable for exceedingly large images. Nevertheless, it has been successfully tested on images ranging from 40x40x40 pixels up to 1500x1500x1500 in 3D and from 40x40 up to approximately 30000x30000 in 2D. The capacity for handling larger images is contingent on the available RAM.

When presented with a new dataset, nnU-Net systematically analyzes the provided training cases to generate a 'dataset fingerprint'. Based on this analysis, it constructs several U-Net configurations tailored to the dataset:

\begin{itemize}
    \item 2D U-Net :- Applicable for both 2D and 3D datasets.
    \item 3D Full Resolution U-Net :-  Operates on high-resolution images and is intended for 3D datasets
    \item 3D Low Resolution U-Net :- Operates on low-resolution images 
    \item 3D Cascade Full Resolution U-Net:- A 3D U-Net cascade where an initial low-resolution 3D U-Net refines predictions through a subsequent high-resolution 3D U-Net. This configuration is applied to large 3D datasets.
\end{itemize}

For datasets with smaller image sizes, the U-Net cascade (and thus the 3D low-resolution configuration) is excluded, as the patch size of the full-resolution U-Net is sufficient to cover a significant portion of the input images. The configuration of nnU-Net's segmentation pipelines is based on a three-step approach:

\begin{itemize}
    \item Fixed Parameters: These parameters remain constant and are not adapted. Through the development of nnU-Net, a robust configuration was identified that includes the loss function, most data augmentation strategies, and the learning rate.
    \item  Rule-Based Parameters: These parameters are adjusted based on the dataset fingerprint using heuristic rules. For instance, network topology, which includes pooling behaviour and network depth, is adapted to the patch size. The patch size, network topology, and batch size are optimized jointly, considering GPU memory constraints.
    \item Empirical Parameters: These parameters are determined through trial and error. This involves selecting the most suitable U-Net configuration for the dataset (2D, 3D full resolution, 3D low resolution, 3D cascade) and optimizing the postprocessing strategy.
\end{itemize}
nnU-Net’s systematic approach to configuring segmentation pipelines based on dataset-specific characteristics and robust default settings makes it a versatile and powerful tool for semantic segmentation tasks.

\subsection{Datasets}

In this study, we utilized five datasets. Those are namely 
Left atrial and Scar Quantification and segmentation Challenge (LAScarQS) 2022 dataset~\citep{zhuang2023left},
2018 Atria Segmentation Challenge (LASC)~\citep{xiong2021global},
Automated Cardiac Diagnosis Challenge (ACDC)-2017~\citep{bernard2018deep},
Multi-Centre, Multi-Vendorand Multi-Disease Cardiac Image Segmentation Challenge (MnM) ~\citep{campello2021multi} and MnM2~\citep{martin2023deep}.

\subsubsection{LAScarQS Challenge Dataset}

The LAScarQS challenge encompasses two primary tasks. The first task involves segmenting the left atrium (LA) cavity and scars, while the second task focuses solely on segmenting the LA cavity. For Task 1, the dataset includes 60 training images with corresponding labels and 10 validation images without labels. Task 2 provides 130 training images with labels and 20 validation images without labels. Consequently, only the training data can be utilized for both training and testing purposes. For Task 1, we allocated 50 images for training and the remaining 10 for testing. For Task 2, we used 115 images for training and 15 for testing.

The LGE-MRIs in this challenge were sourced from the University of Utah, Beth Israel Deaconess Medical Center, and King's College London. The scans were performed using Siemens Avanto 1.5 T, Siemens Vario 3 T, or Philips Acheiva 1.5 T MRI machines. Scans were acquired either free-breathing with navigator-gating or using navigator-gating with fat suppression. The spatial resolution of the scans varied: 1.25 × 1.25 × 2.5 mm, 1.4 × 1.4 × 1.4 mm, or 1.3 × 1.3 × 4.0 mm. Patients underwent MRI scans either before undergoing ablation or between one and six months post-ablation.

\subsubsection{2018 Left Atria Segmentation Challenge (LASC) Dataset}

The 2018 Left Atria Segmentation Challenge (LASC) concentrated on the segmentation of the LA cavity. The dataset included 100 training images and 54 testing images, all provided with 3D binary masks of the LA cavity. Each LGE-MRI scan featured a spatial resolution of 0.625 × 0.625 × 0.625 mm$^3$, with spatial dimensions of either 576 × 576 × 88 or 640 × 640 × 88 pixels. These clinical images were obtained using either a 1.5 Tesla Avanto or a 3.0 Tesla Verio whole-body scanner (Siemens Medical Solutions, Erlangen, Germany). The LA cavity volumes were meticulously segmented in consensus and agreement by three trained observers, ensuring the provision of high-quality ground truth annotations for both training and evaluation.

\subsubsection{Automated Cardiac Diagnosis Challenge (ACDC) 2017 Dataset}

The Automated Cardiac Diagnosis Challenge (ACDC) 2017 dataset comprises 150 MRI scans categorized into five subgroups: normal, previous myocardial infarction, dilated cardiomyopathy, hypertrophic cardiomyopathy, and abnormal right ventricle. These scans were collected over six years using two MRI scanners with different magnetic strengths: 1.5 Tesla (Siemens Area, Siemens Medical Solutions, Germany) and 3.0 Tesla (Siemens Trio Tim, Siemens Medical Solutions, Germany). Cine MRI images were acquired under breath-hold conditions using either retrospective or prospective gating, with a steady-state free precession (SSFP) sequence in the short-axis orientation. The scans consist of a series of short-axis slices covering the left ventricle (LV) from base to apex, with a slice thickness of 5 mm (occasionally 8 mm) and sometimes an interslice gap of 5 mm, resulting in images spaced every 5 or 10 mm depending on the examination. The spatial resolution ranges from 1.37 to 1.68 mm$^2$/pixel, and each series includes 28 to 40 images, covering the cardiac cycle completely or partially. The dataset is divided into 100 training images and 50 testing images for the segmentation of the left ventricle (LV), myocardium (MYO), and right ventricle (RV) during both end-systolic (ES) and end-diastolic (ED) phases.

\subsubsection{Multi-Centre, Multi-Vendor, and Multi-Disease Cardiac Image Segmentation Challenge (MnM-1 and MnM-2)}

The MnM challenge has been conducted twice, first in 2020 (MnM-1) and then in 2021 (MnM-2). MnM-1 included a total of 345 scans, with 209 images designated for training and 136 for testing. Participants were tasked with segmenting the left ventricle (LV), myocardium (MYO), and right ventricle (RV) in both end-systolic (ES) and end-diastolic (ED) phases. The scans were obtained from clinical centres located in three countries (Spain, Germany, and Canada) and utilized four different magnetic resonance scanner vendors: Siemens, General Electric, Philips, and Canon.

\begin{figure}[t]
    \centering
    \includegraphics[width=\linewidth]{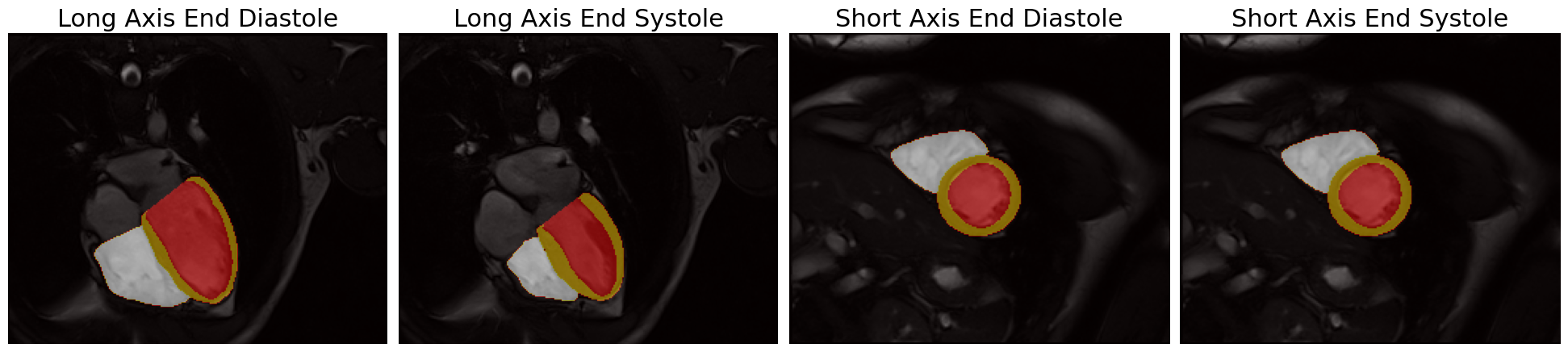}
    \caption{Visualization of the long axis and short axis views in both end diastole and end systole phases for the MnM2 dataset. The right ventricle (RV) is highlighted in white, the Myocardium (MYO) is highlighted in yellow, and the Left Ventricle (LV) is highlighted in red. }
    \label{fig:mnm2_view}
\end{figure}

MnM-2 provided a training set of 200 images and a testing set of 160 images. Similar to MnM-1, segmentation was required for the LV, MYO, and RV in both ES and ED phases. However, MnM-2 included both Short-Axis (ShA) and long-axis (LoA) views. The LoA view shows the heart from base to apex, essentially cutting the heart vertically, while the ShA view cuts the heart horizontally, perpendicular to the long axis. It shows circular cross-sections of the ventricles. Fig.~\ref{fig:mnm2_view} shows the LoA and ShA for both ES and ED phases. The data for MnM-2 were acquired from clinical centres in Spain using three different MRI scanner vendors: Siemens, General Electric, and Philips.

A summary of the datasets, including labels and the number of training and testing images, are shown in supplementary material Table~\ref{tab:sup_summary_cardiac_datasets}.

\subsection{Implementation Details}

In this study, we employed nnU-Net, which supports training under five main conditions: 2D, 3D full resolution, 3D low resolution, 3D cascade, and ensemble. However, it was not feasible to evaluate certain datasets using the 3D low-resolution and cascade configurations. For datasets with small image sizes, the U-Net cascade (and consequently the 3D low-resolution configuration) was omitted because the patch size of the full-resolution U-Net already covered a substantial portion of the input images.

The models were trained using an NVIDIA A100 80GB PCIe GPU. The training process spanned 1000 epochs, starting with an initial learning rate of 0.01. We utilised the Stochastic Gradient Descent (SGD) optimiser. To ensure the robustness and reliability of the model's performance, we used five-fold cross-validation and the test results were obtained using all the five folds as explained in the nnU-Net.

\subsection{Evaluation Metrics}

To assess the performance of our segmentation models, we employ a comprehensive set of evaluation metrics: Dice Similarity Coefficient (DSC), Jaccard Index, Hausdorff Distance (HD), Mean Surface Distance (MSD), and the 95th percentile Hausdorff Distance (HD95). Each of these metrics provides unique insights into different aspects of the segmentation quality, offering a holistic view of model performance.

\section{Results}
Here, we present the results of the nnU-Net models on the LAScarQS, LASC, ACDC, MnM, and MnM2 datasets. Note that when comparing with other approaches, the results are reported with the performance of the best-performing method shown first, followed by nnU-Net's best performance (i.e., the best-performing method vs. nnU-Net's best performance).

\subsection{LAScarQS}

\begin{table}[t]
\setlength{\tabcolsep}{3pt}
\caption{Performance of LAScarQS (Task 1). The best cavity segmentation values are in \textbf{bold}, and the best scar segmentation values are \underline{underlined}. DSC- Dice Score, HD - Hausdorff Distance, MSD- Mean Surface Distance, HD95-95th percentile of HD.}
\label{tab:lascar1_nn_only}
\centering
\begin{tabular}{llccccc}
\toprule
Model & Label  & DSC    & Jaccard & HD     & MSD    & HD95   \\ \midrule
\multirow{2}{*}{2D}     & Cavity & 0.926  & 0.863   & 12.952 & 0.805  & 3.402  \\
                        & Scar   & 0.438  & 0.283   & 37.166 & 2.539  & 13.036 \\ \midrule
\multirow{2}{*}{\parbox[t]{1.6cm}{3D full\\ resolution}} & Cavity & 0.939  & 0.884   & 12.622 & 0.666  & 3.088  \\
 & Scar   & 0.443  & 0.288   & \underline{37.060} & \underline{2.512}  & 12.620 \\ \midrule
\multirow{2}{*}{\parbox[t]{1.6cm}{3D low\\ resolution}}  & Cavity & 0.937  & 0.882   & 13.942 & 0.711  & 3.254  \\
  & Scar   & 0.411  & 0.262   & 37.294 & 2.789  & 13.425 \\ \midrule
\multirow{2}{*}{\parbox[t]{1.6cm}{3D cascade}}  & Cavity & \textbf{0.939}  & 0.885   & 12.601 & 0.674  & 3.138  \\
  & Scar   & \underline{0.449}  & \underline{0.293}   & 38.125 & {2.530}  & \underline{12.554} \\ \midrule
\multirow{2}{*}{Ensemble}   & Cavity & \textbf{0.939}  & \textbf{0.886}   & \textbf{12.486} & \textbf{0.663}  & \textbf{3.041}  \\
 & Scar   & 0.439  & 0.285   & 37.078 & 2.590  & 12.850 \\ \bottomrule
\end{tabular}
\end{table}

\begin{figure*}[t]
    \centering
    \begin{subfigure}{0.3\textwidth}
        \includegraphics[width=\textwidth]{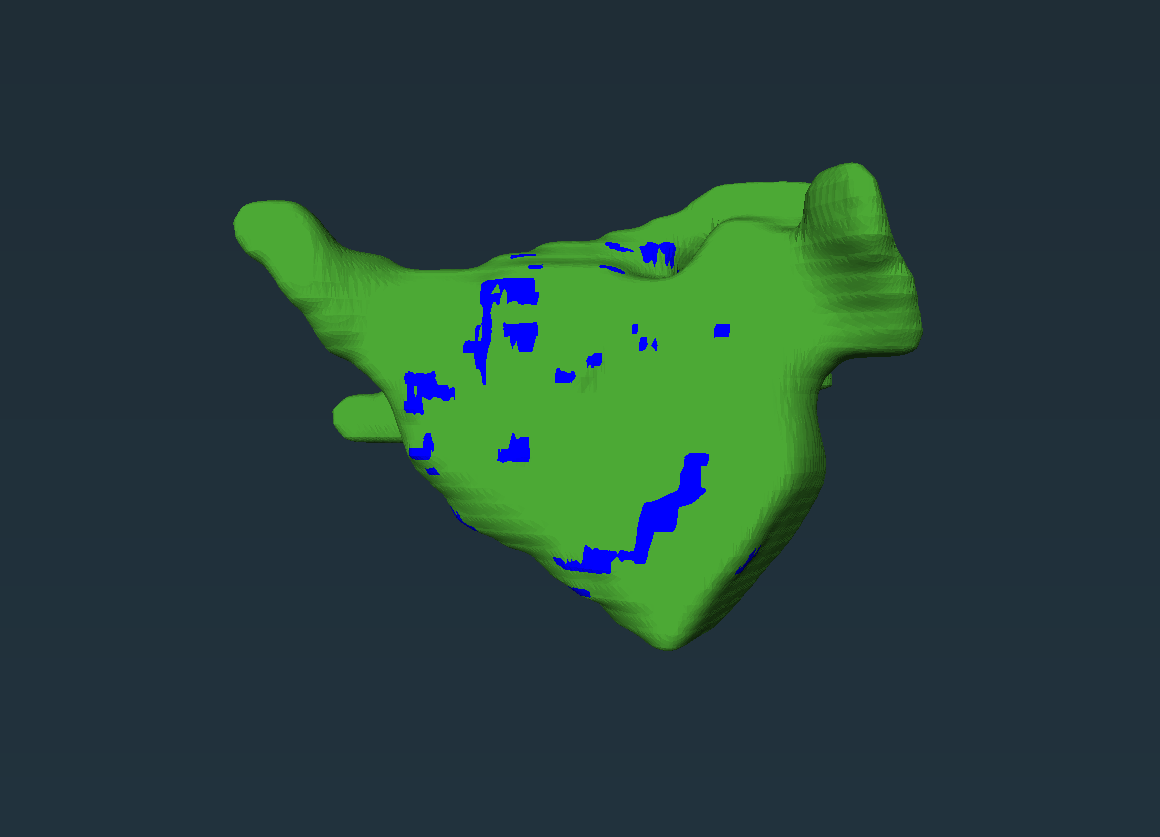}
        \caption{Ground truth}
        \label{fig:1}
    \end{subfigure}
    \hfill
    \begin{subfigure}{0.3\textwidth}
        \includegraphics[width=\textwidth]{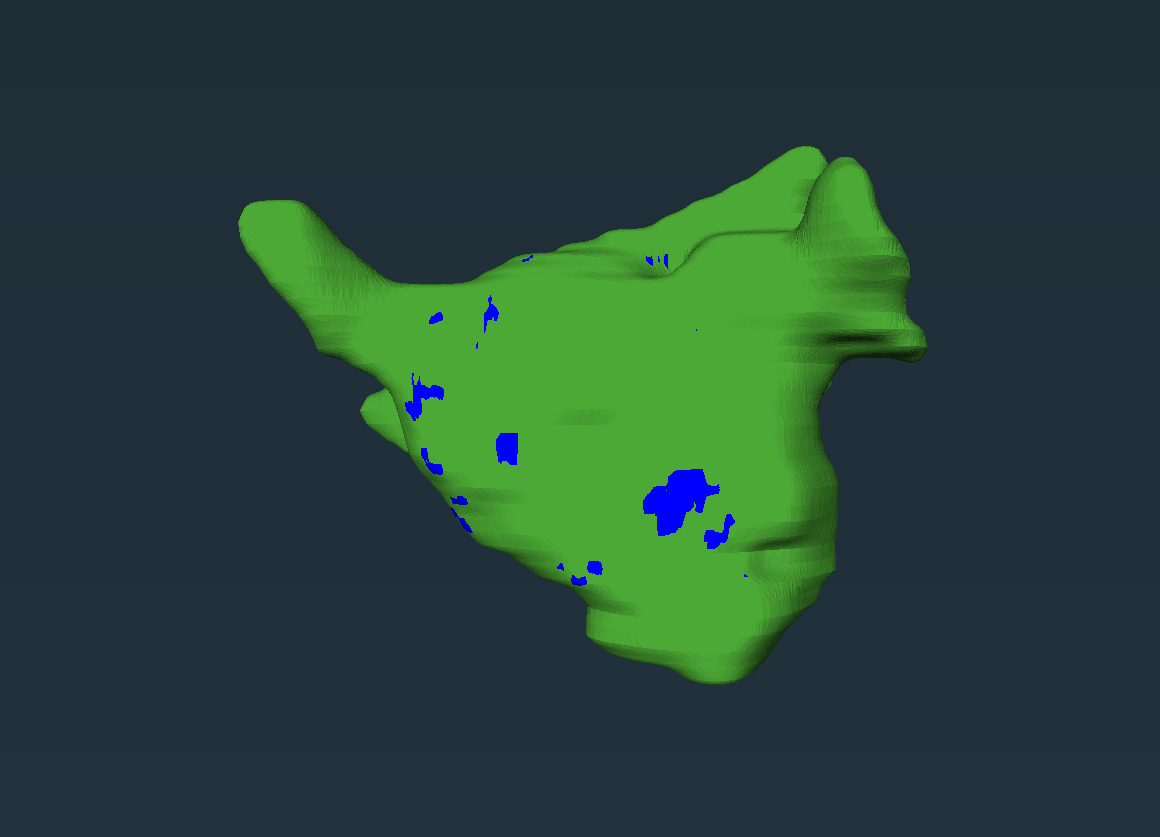}
        \caption{2D nnU-Net}
        \label{fig:2}
    \end{subfigure}
    \hfill
    \begin{subfigure}{0.3\textwidth}
        \includegraphics[width=\textwidth]{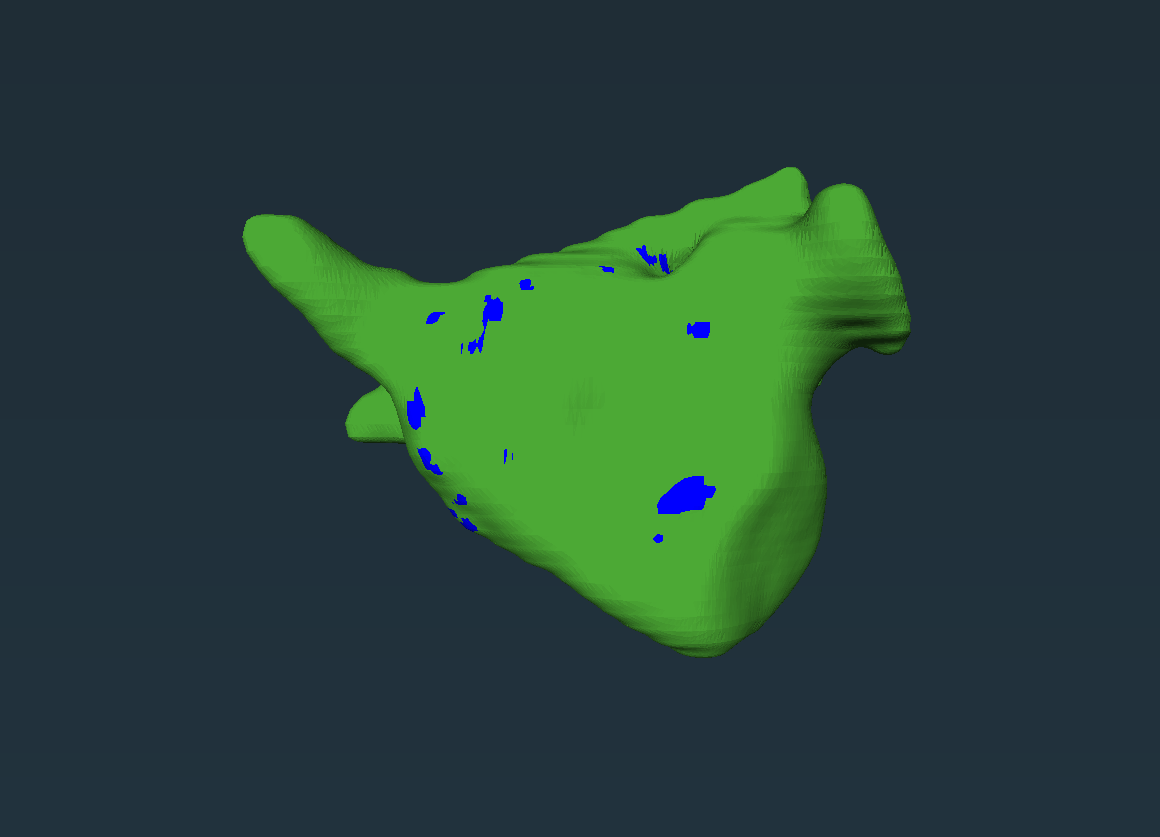}
        \caption{3D full resolution}
        \label{fig:3}
    \end{subfigure}

    \vskip\baselineskip
    \vspace{-2em}

    \begin{subfigure}{0.3\textwidth}
        \includegraphics[width=\textwidth]{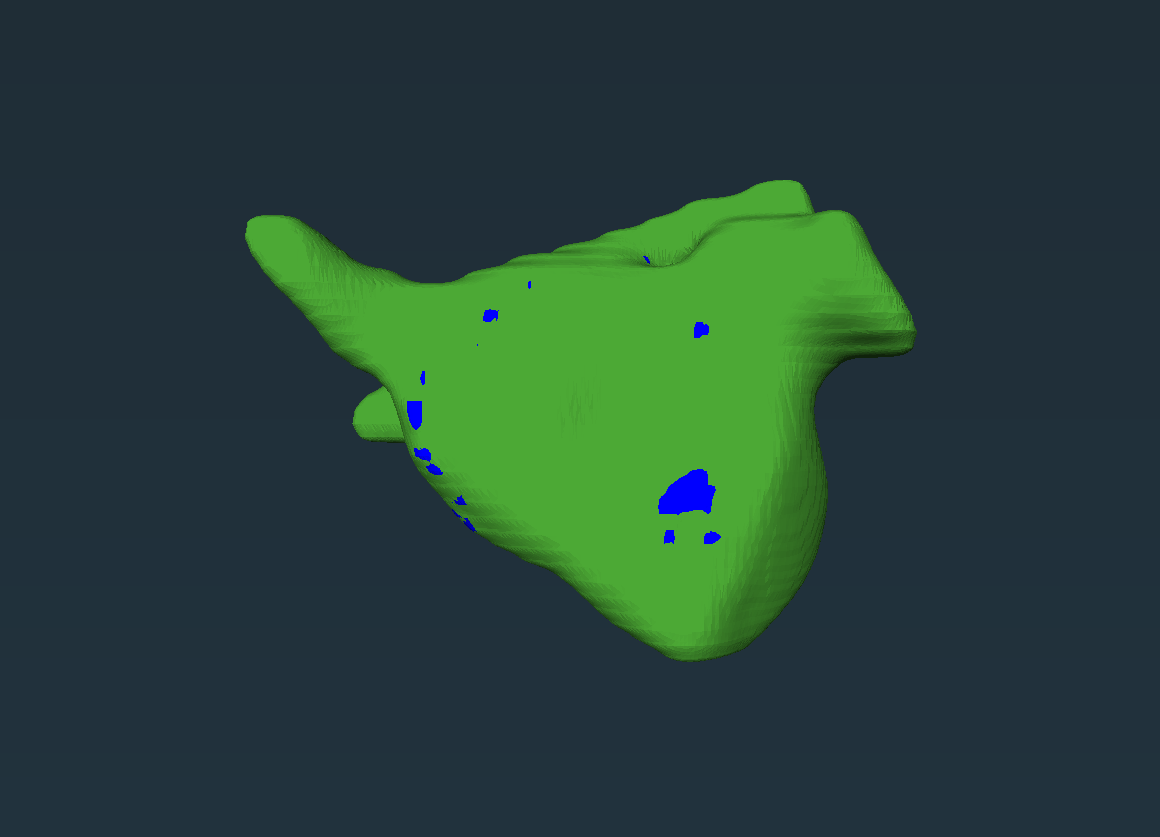}
        \caption{3D low resolution}
        \label{fig:4}
    \end{subfigure}
    \hfill
    \begin{subfigure}{0.3\textwidth}
        \includegraphics[width=\textwidth]{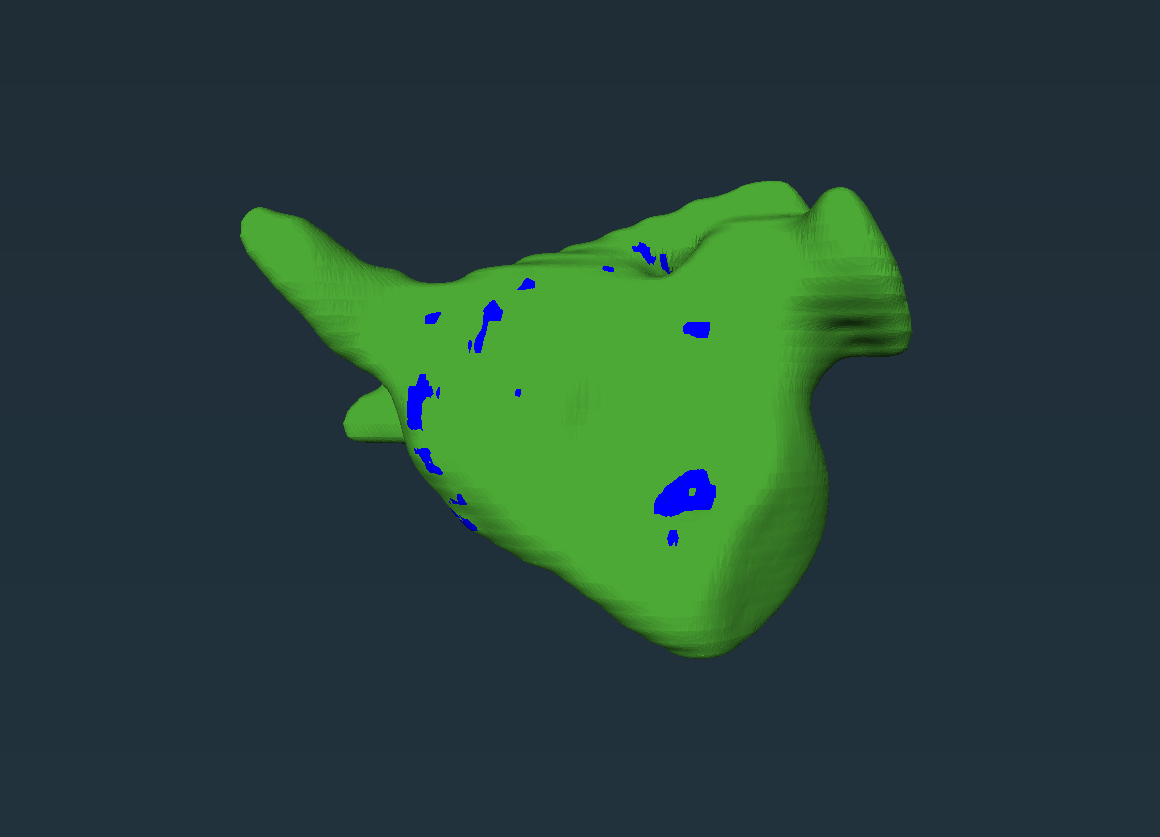}
        \caption{3D cascade}
        \label{fig:5}
    \end{subfigure}
    \hfill
    \begin{subfigure}{0.3\textwidth}
        \includegraphics[width=\textwidth]{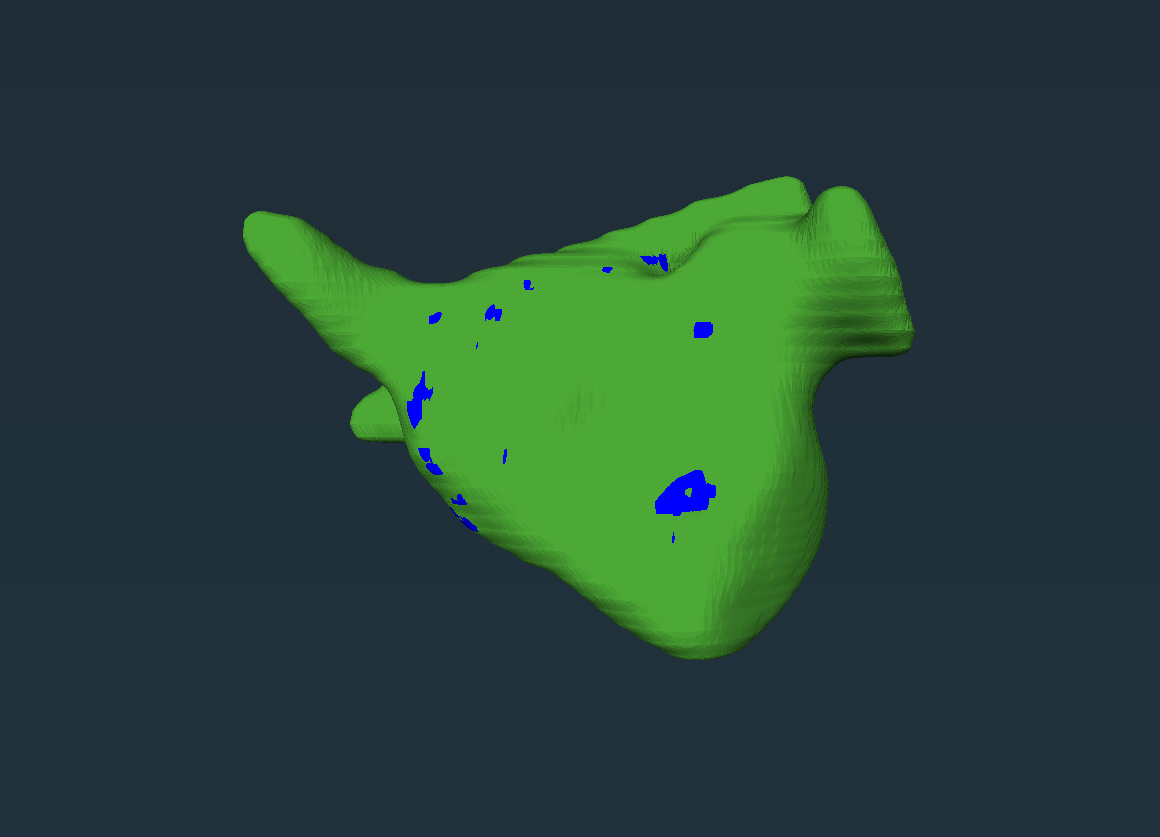}
        \caption{Ensemble}
        \label{fig:6}
    \end{subfigure}
\caption{Comparison of Ground Truth and Predictions from different variations of nnU-Nets for the LAScarQs Task 1. The Left Atrial (LA) cavity is highlighted in green, and LA scars are highlighted in blue. Visualised using Amira 3D software.}
\vspace{-1.5em}
    \label{fig:lascarqs}
\end{figure*}
% \vspace{-2em}
\begin{table}[h]
\setlength{\tabcolsep}{3pt}
\caption{Performance of LAScarQS (Task 2). The best cavity segmentation values are in \textbf{bold}.  DSC- Dice Score, HD - Hausdorff Distance, MSD- Mean Surface Distance, HD95-95th percentile of HD}
\label{tab:lascar2_nn_only}
\centering
\begin{tabular}{lccccc}
\toprule
Model        & DSC    & Jaccard & HD     & MSD    & HD95   \\ \midrule
2D           & 0.930  & 0.869   & 13.971 & 0.733  & 3.018  \\
3D full resolution & 0.937  & 0.882   & 12.971 & 0.672  & 2.880  \\
3D low resolution   & 0.935  & 0.879   & \textbf{12.741} & 0.692  & 3.069  \\
3D cascade   & 0.937  & 0.882   & 12.807 & 0.667  & 2.746  \\
Ensemble     & \textbf{0.938}  & \textbf{0.883}   & 12.767 & \textbf{0.652}  & \textbf{2.737} \\ \bottomrule
\end{tabular}
\end{table}

% In Table~\ref{tab:lascar1_nn_only} and Table~\ref{tab:lascar2_nn_only}, we compare the performance of the nnU-net in Task 1 and Task 2 respectively. In Table~\ref{tab:sup_lascar1_compare} and Table~\ref{tab:sup_lascar2_compare} in supplementary materials, we compare the results against other methods. One notable observation is that, DSC and Jaccard values of the scar are low compared to the cavity segmentation. Also,  HD, MSD and HD95 values are much higher. Cavity segmentation in both Task 1 and Task 2, the nnU-Net ensemble model achieves the best performance in every metrics, while the nnU-Net (3D low res) model achieves the best performance for HD in Taks 2. nnU-Net is able to perform competitively even with lesser data compared to other methods in the challenge. Fig.~\ref{fig:lascarqs} provides a qualitative comparison of LAScarQS Task1 performance, visualized using Amira 3D software \citep{stalling2005amira}.
In Table~\ref{tab:lascar1_nn_only} and Table~\ref{tab:lascar2_nn_only}, we present the performance evaluation of the nnU-Net for Task 1 and Task 2, respectively. Comparative analyses against alternative methods are provided in the supplementary materials, in Table~\ref{tab:sup_lascar1_compare} and Table~\ref{tab:sup_lascar2_compare}. When comparing our methods to others, the LAScarQS Task 1 scar segmentation exhibited the most significant difference, with other methods (DSC - 0.660 to 0.553 vs 0.439 to 0.411) surpassing the nnU-Net models by 21.1\%. Additionally, the HD values for scar segmentation are notably higher (Table \ref{tab:lascar1_nn_only} and supplementary Table~\ref{tab:sup_lascar1_compare}). This might be due to the class imbalance of the scars and cavity (see discussion section for more details).

However, nnU-Net models achieve superior performance in cavity segmentation, despite their lower results in scar segmentation in Task 1 (DSC - 0.875 to 0.938 vs 0.926 to 0.939). This trend is also observed in LAScarQS Task 2 cavity segmentation (Table \ref{tab:lascar2_nn_only} and supplementary Table~\ref{tab:sup_lascar2_compare}), where even the nnU-Net (2D) model outperforms other methods (DSC range - 0.872 to 0.929 vs 0.930 to 0.938). In Task 2, the nnU-Net ensemble model achieves the best performance in both Dice score (0.938) and MSD (0.652) metrics, while the nnU-Net (3D low res) model achieves the best performance for HD (12.741). nnU-Net is able to perform competitively even with lesser data compared to other methods in the challenge. The nnU-Net models achieve higher performance metrics not only in dice scores but also in HD and MSD matrices.

\subsection{LASC}
% \vspace{-1em}
\begin{table}[t]
\setlength{\tabcolsep}{3pt}
\caption{Performance of LASC dataset. The best cavity segmentation values are in \textbf{bold}.  DSC- Dice Score, HD - Hausdorff Distance, MSD- Mean Surface Distance, HD95-95th percentile of HD}
\label{tab:lasc_nn_only}
\centering
\begin{tabular}{lccccc}
\toprule
Model        & DSC    & Jaccard & HD     & MSD    & HD95   \\ \midrule
2D           & 0.926  & 0.863   & 17.583 & 1.052  & 3.930  \\
3D full resolution & 0.933  & 0.875   & 17.485 & 0.972  & 3.681  \\
3D low resolution   & 0.931  & 0.872   & 16.877 & 0.991  & 3.727  \\
3D cascade   & 0.933  & 0.874   & 17.553 & 0.984  & 3.756  \\
Ensemble     & \textbf{0.934}  & \textbf{0.877}   & \textbf{16.873} & \textbf{0.954}  & \textbf{3.628} \\ \bottomrule
\end{tabular}
\vspace{-1.2em}
\end{table}

\begin{figure*}[t]
    \centering
    \includegraphics[width=\textwidth]{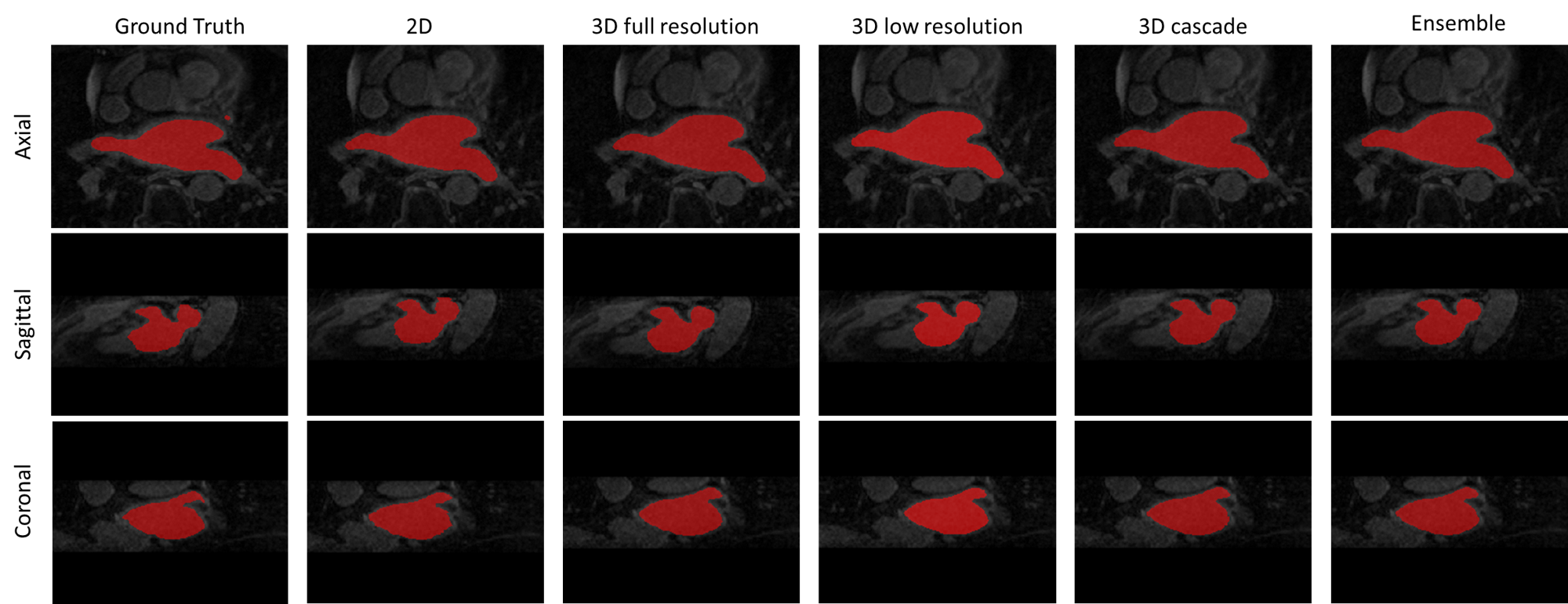}
\caption{Comparison of ground truth and predictions from different nnU-Net Versions (2D, 3D Full Resolution, 3D Low Resolution, 3D Cascade, and Ensemble) in three anatomical views: Axial, Sagittal, and Coronal for the LASC dataset. The cavity area is highlighted in Red. Visualized using ITK-SNAP software.}
    \label{fig:lasc}
\end{figure*}

For the LASC dataset, the ensemble model achieves the highest performance (DSC - 0.934) (Table~\ref{tab:lasc_nn_only}). According to supplementary Table~\ref{tab:sup_LASC_compare}, nnU-Net demonstrates competitive performance with other methods, with only~\cite{singh2023attention} surpassing nnU-Net by 0.1\% (DSC range - 0.898 to 0.935 vs 0.923 to 0.934). Interestingly, even the nnU-Net (2D) model shows competitive performance compared to the latest models~\citep{xu2024dynamic} (DSC in both 0.926). nnU-Net is able to surpass the novel method even without additional configurations. We assess the qualitative performance of the nnU-Nets using ITK-SNAP software~\citep{yushkevich2016itk} as shown in Fig.~\ref{fig:lasc} for axial, sagittal and coronal views.

\subsection{ACDC}
\begin{table}[h]
\setlength{\tabcolsep}{3pt}
\caption{Performance of ACDC Dataset for End-Diastole (ED) and End-Systole (ES) phases for 2D, 3D full resolution (3D full.) and Ensemble (Ens.) models. The best values for the Right Ventricle (RV), Myocardium (MYO), and Left Ventricle (LV) are highlighted in \textbf{bold}, \underline{underline}, and \textit{italic}, respectively.}
\label{tab:acdc_corrected}
\centering
\begin{tabular}{llcccccccccc}
\toprule
                 &       & \multicolumn{5}{c}{End-Diastole (ED) phase} & \multicolumn{5}{c}{End-Systole (ES) phase} \\
\cmidrule(lr){3-7} \cmidrule(lr){8-12}
Model            & Label & DSC    & Jaccard & HD      & MSD    & HD95   & DSC    & Jaccard & HD      & MSD    & HD95   \\
\midrule
\multirow{3}{*}{2D}               & RV    & 0.942  & 0.892   & \textbf{10.438} & 0.467  & 3.152  & 0.885  & 0.799   & 12.678 & 0.827  & 4.318  \\
               & MYO   & 0.897  & 0.814   & 10.05  & 0.331  & \underline{1.583}  & 0.913  & 0.841   & \underline{8.231}  & \underline{0.384}  & \underline{1.975}  \\
               & LV    & \textit{0.965}  & \textit{0.933}   & \textit{6.739}  & \textit{0.347}  & \textit{2.350}  & 0.927  & 0.868   & \textit{6.795}  & \textit{0.483}  & \textit{2.713}  \\

               \midrule

\multirow{3}{*}{\parbox[t]{1.2cm}{3D full\\ resolution}}.& RV    & 0.934  & 0.880   & 11.494 & 0.617  & 3.861  & 0.882  & 0.793   & 12.743 & 0.911  & 5.245  \\
& MYO   & 0.889  & 0.801   & \underline{8.057}  & 0.367  & 2.135  & 0.906  & 0.829   & 8.785  & 0.456  & 2.551  \\
& LV    & 0.959  & 0.922   & 8.486  & 0.443  & 2.720  & 0.901  & 0.831   & 9.028  & 0.972  & 4.968  \\ \midrule
\multirow{3}{*}{\parbox[t]{1.2cm}{Ensemble}}& RV    & \textbf{0.944}  & \textbf{0.896}   & 10.716 & \textbf{0.459}  & \textbf{3.110}  & \textbf{0.892}  & \textbf{0.809}   & \textbf{12.200} & \textbf{0.751}  & \textbf{4.208}  \\
& MYO   & \underline{0.898}  & \underline{0.816}   & 9.884  & \underline{0.325}  & 1.818  & \underline{0.915}  & \underline{0.844}   & 8.460  & 0.384  & 2.193  \\
& LV    & 0.963  & 0.930   & 9.584  & 0.404  & 2.474  & \textit{0.922}  & \textit{0.861}   & 8.321  & 0.608  & 3.477  \\
\bottomrule
\end{tabular}
\vspace{-1.2em}
\end{table}

\begin{figure*}[!h]
    \centering
    \includegraphics[width=\linewidth]{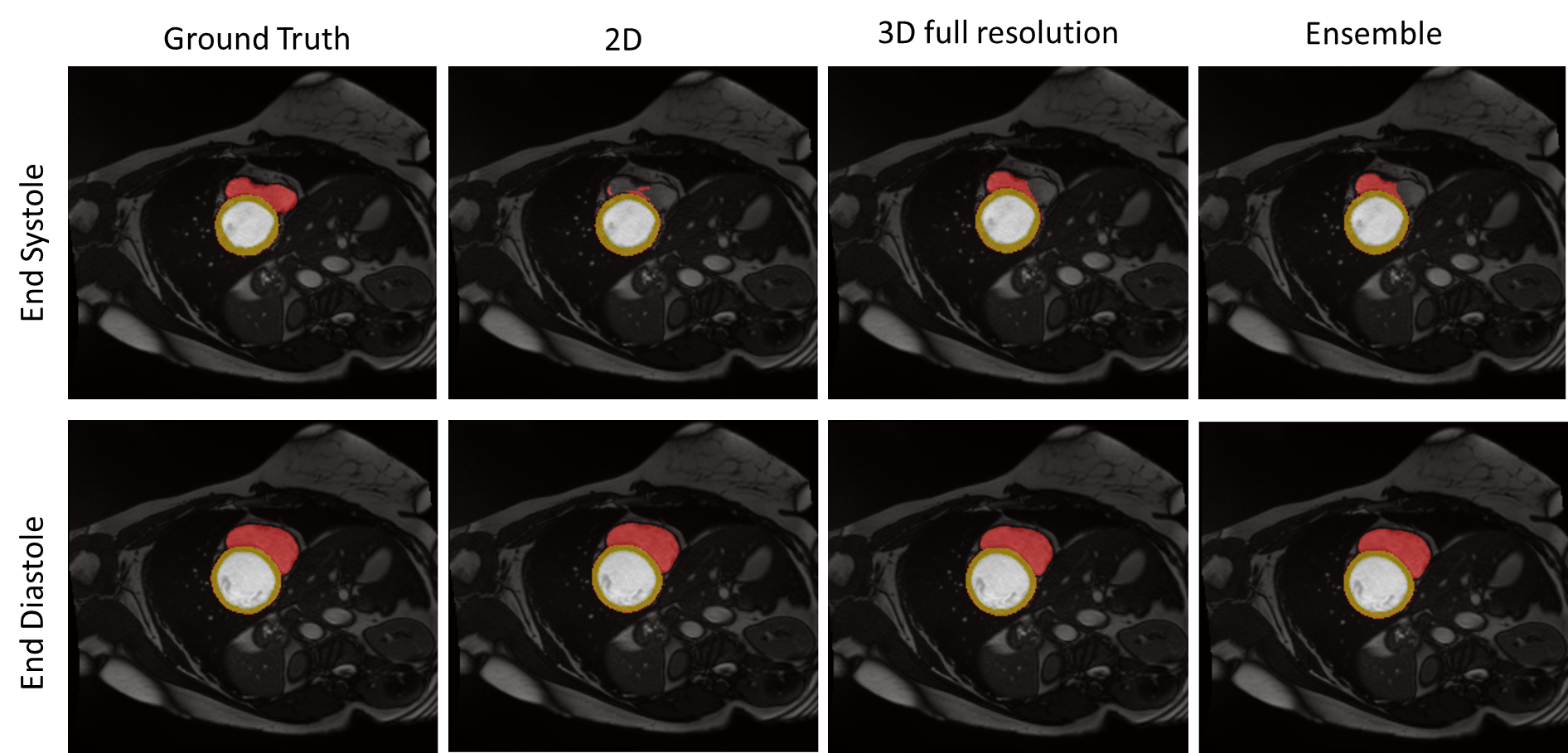}
    \caption{Comparison of Ground Truth and Predictions from nnU-Net Variants (2D, 3D Full Resolution, and Ensemble) on the ACDC Dataset for End Systole (ES) and End Diastole (ED) Phases. The right ventricle (RV) is highlighted in red, the myocardium (MYO) is in yellow, and the left ventricle (LV) is in white.}
    \label{fig:acdc}
\end{figure*}

Performance evaluation of the ACDC dataset is conducted under two main conditions: End-Diastole (ED)  and End-Systole (ES) (Table~\ref{tab:acdc_corrected}). In both cases, overall,  the ensemble method demonstrates superior performance compared to other variations of nnU-Nets. Surprisingly, the 2D nnU-Net exhibits better performance in DSC (0.965) than both 3D (0.959) and ensemble (0.963) models in LV segmentation of the ACDC-ED phase. When comparing the ED and ES phases, LV and RV generally perform better in the ED phase than in the ES phase. However, MYO has shown better performance in the ES phase. 

When compared to other approaches (Supplementary Table~\ref{tab:sup_acdc_compare}), nnU-Net exhibits slightly lower performance across all segmentation tasks relative to other methods tailored to the ACDC dataset. However, the differences in Dice Similarity Coefficient (DSC) are consistently less than 2\% in all cases: LV ED (0.968 vs. 0.965), LV ES (0.938 vs. 0.927), MYO ED (0.906 vs. 0.898), MYO ES (0.923 vs. 0.915), RV ED (0.955 vs. 0.944), and RV ES (0.904 vs. 0.892). Fig.~\ref{fig:acdc} illustrates a performance comparison between the ground truth and predictions generated by nnU-Net (2D), nnU-Net (3D full resolution), and ensemble models for both ED and ES phases.

\vspace{-1.2em}
\begin{table}[!htb]
\setlength{\tabcolsep}{3pt}
\caption{Performance of \textbf{MnM} Dataset for End-Diastole (ED) and End-Systole (ES) phases for 2D, 3D full resolution (3D full.) and Ensemble (Ens.). The best segmentation values for the Left Ventricle (LV), Myocardium (MYO), and Right Ventricle (RV) are highlighted in \textbf{bold}, \underline{underline}, and \textit{italic}, respectively. DSC- Dice Score, HD - Hausdorff Distance, MSD- Mean Surface Distance, HD95-95th percentile of HD, LV- Left Ventricle, MYO- Myocardium, RV- Right Ventricle.}
\label{tab:MnM_merged}
\centering
\begin{tabular}{llcccccccccc}
\toprule
                 &       & \multicolumn{5}{c}{End-Diastole (ED) phase} & \multicolumn{5}{c}{End-Systole (ES) phase} \\
\cmidrule(lr){3-7} \cmidrule(lr){8-12}
Model            & Label & DSC    & Jaccard & HD      & MSD    & HD95   & DSC    & Jaccard & HD      & MSD    & HD95   \\
\midrule
\multirow{3}{*}{2D}                & LV    & 0.936  & 0.882   & 7.517  & 0.728  & 3.871  & 0.888  & 0.833   & 12.681 & 2.368  & 8.513  \\
              & MYO   & 0.824  & 0.706   & 10.738 & 0.592  & 3.676  & 0.800  & 0.689   & 15.428 & 1.905  & 7.304  \\
               & RV    & 0.909  & 0.836   & 11.601 & 0.900  & 4.578  & \textit{0.893}  & \textit{0.821}   & 14.595 & 1.450  & 6.411  \\ \midrule
               
\multirow{3}{*}{\parbox[t]{1.2cm}{3D full\\ resolution}}     & LV    & 0.933  & 0.877   & 8.199  & 0.819  & 4.261  & \textbf{0.909}  & \textbf{0.842}   & 8.576  & \textbf{0.944}  & 4.507  \\
     & MYO   & 0.819  & 0.699   & 10.776 & 0.580  & 3.484  & 0.841  & 0.734   & 11.141 & 0.812  & 3.952  \\
     & RV    & 0.908  & 0.836   & 11.520 & 0.870  & 4.501  & 0.871  & 0.784   & 13.130 & 1.258  & 5.673  \\ \midrule
\multirow{3}{*}{Ensemble}     & LV    & \textbf{0.937}  & \textbf{0.883}   & \textbf{7.393}  & \textbf{0.725}  & \textbf{3.761}  & 0.888  & 0.803   & \textbf{8.486}  & 0.956  & \textbf{4.432}  \\
        & MYO   & \underline{0.826}  & \underline{0.709}   & \underline{9.944}  & \underline{0.527}  & \underline{3.138}  & \underline{0.864}  & \underline{0.762}   & \underline{9.872}  & \underline{0.613}  & \underline{3.542}  \\
         & RV    & \textit{0.913}  & \textit{0.843}   & \textit{10.847} & \textit{0.818}  & \textit{4.208}  & 0.852  & 0.751   & \textit{12.658} & \textit{1.083}  & \textit{5.366}  \\
\bottomrule
\end{tabular}
\end{table}

\subsection{MnM}

\begin{figure}[!htb]
    \centering
    \includegraphics[width=\textwidth]{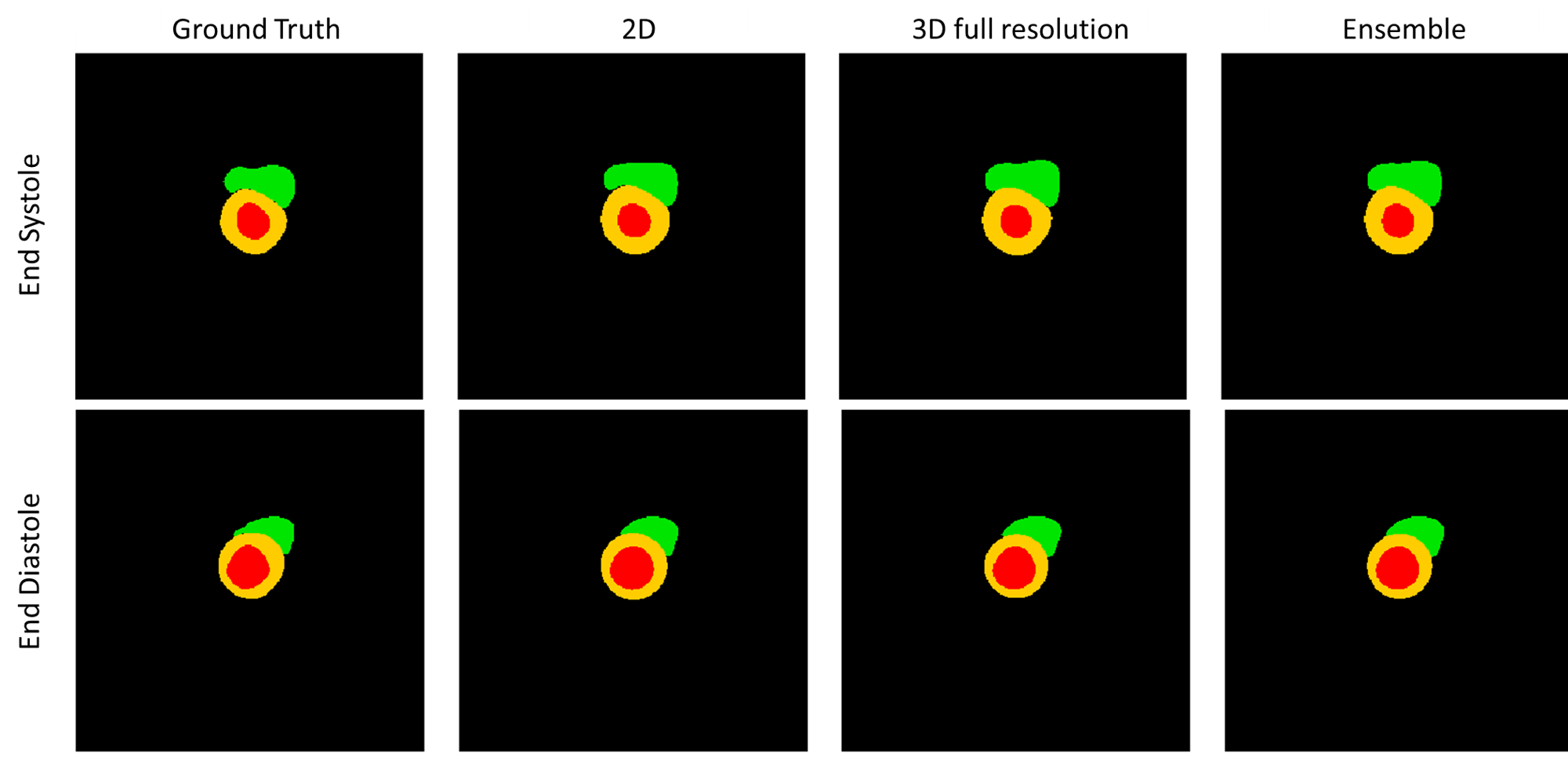}
\caption{Comparison of Ground Truth and Predictions from nnU-Net Variants (2D, 3D Full Resolution, and Ensemble) on the MnM Dataset for End Systole (ES) and End Diastole (ED) Phases. The right ventricle (RV) is highlighted in green, the myocardium (MYO) is in yellow, and the left ventricle (LV) is in red.}
    \label{fig:mnm_es}
\end{figure}

As in the ACDC dataset, MnM performance is evaluated on both ES and ED (Table~\ref{tab:MnM_merged}) phases. The 2D nnU-Net (DSC-0.893) outperforms both 3D (DSC-0.871) and ensemble models (DSC-0.852) in RV segmentation in the ES phase, while the 3D full-resolution model also outperforms LV segmentation in terms of dice score in the ES phase (0.888 vs 0.909). In the ED phase, the ensemble model demonstrates superior performance (DSC - 0.937, HD 7.4). RV segmentation in both phases achieves higher dice scores (ED - 0.552 to 0.910 vs 0.908 to 0.913, ES - 0.517 to 0.860 vs 0.852 to 0.871) compared to other approaches (supplementary Table~\ref{tab:sup_mnm1_compare}). In other cases, other approaches surpass the nnU-Net by slight margins, typically less than 1\% in DSC (LV (ED)-0.940 vs 0.937, MYO (ED) - 0.839 to 0.826, MYO (ES) - 0.870-0.864). Similar to the ACDC dataset, MnM also shows better performance in LV and RV in the ED phase especially in DSC and Jaccard values compared to the ES phase. However, MYO has shown better performance in the ES phase in generally.In Fig.~\ref{fig:mnm_es}, we compare the performance of the nnU-Net models in both ES and ED phases.

\subsection{MnM2}

\begin{table}[!h]
\setlength{\tabcolsep}{3pt}
\caption{Performance of \textbf{MnM2} Dataset for Short Axis (ShA) End-Diastole (ED) and End-Systole (ES) phases for 2D, 3D full resolution (3D full.) and Ensemble (Ens.) . The best segmentation values for the Left Ventricle (LV), Myocardium (MYO), and Right Ventricle (RV) are highlighted in \textbf{bold}, \underline{underline}, and \textit{italic}, respectively. DSC- Dice Score, HD - Hausdorff Distance, MSD- Mean Surface Distance, HD95-95th percentile of HD, LV- Left Ventricle, MYO- Myocardium, RV- Right Ventricle.}
\label{tab:mnm2_SA_merged}
\centering
\begin{tabular}{llcccccccccc}
\toprule
                 &       & \multicolumn{5}{c}{End-Diastole (ED) phase} & \multicolumn{5}{c}{End-Systole (ES) phase} \\
\cmidrule(lr){3-7} \cmidrule(lr){8-12}
Model            & Label & DSC    & Jaccard & HD      & MSD    & HD95   & DSC    & Jaccard & HD      & MSD    & HD95   \\
\midrule
\multirow{3}{*}{2D}                & LV    & 0.957  & 0.920   & 8.268  & 0.515  & \textbf{3.170}  & 0.958  & 0.920   & 8.350  & 0.513  & 3.170  \\
               & MYO   & 0.867  & 0.769   & 12.238 & 0.442  & 2.842  & 0.867  & 0.770   & 11.928 & 0.428  & 2.571  \\
               & RV    & 0.934  & 0.879   & \textit{10.050} & 0.766  & 4.084  & 0.934  & 0.879   & 11.228 & 0.817  & 4.520  \\ \midrule
\multirow{3}{*}{\parbox[t]{1.2cm}{3D full\\ resolution}}     & LV    & 0.955  & 0.916   & 8.361  & 0.565  & 3.571  & 0.956  & 0.916   & 8.233  & 0.561  & 3.481  \\
     & MYO   & 0.862  & 0.761   & 12.035 & 0.426  & 2.561  & 0.862  & 0.761   & 12.004 & 0.426  & 2.555  \\
    & RV    & 0.934  & 0.878   & 10.394 & 0.779  & 4.200  & 0.934  & 0.878   & \textit{10.301} & 0.779  & 4.302  \\ \midrule
\multirow{3}{*}{Ensemble}         & LV    & \textbf{0.958}  & \textbf{0.921}   & \textbf{8.029}  & \textbf{0.496}  & 3.256  & \textbf{0.958}  & \textbf{0.920}   & \textbf{8.225}  & \textbf{0.503}  & \textbf{3.264}  \\
         & MYO   & \underline{0.869}  & \underline{0.772}   & \underline{11.492} & \underline{0.396}  & \underline{2.371}  & \underline{0.868}  & \underline{0.771}   & \underline{11.684} & \underline{0.398}  & \underline{2.341}  \\
        & RV    & \textit{0.937}  & \textit{0.884}   & 11.079 & \textit{0.742}  & \textit{4.021}  & \textit{0.938}  & \textit{0.885}   & 11.119 & \textit{0.722}  & \textit{3.930}  \\
\bottomrule
\end{tabular}
\end{table}

\begin{table}[!h]
\setlength{\tabcolsep}{3pt}
\caption{Performance of \textbf{MnM2} Dataset for Long Axis (LoA) End-Diastole (ED) and End-Systole (ES) phases for 3D full resolution only.}
\label{tab:mnm2_LA_merged}
\centering
\begin{tabular}{lcccccccccc}
\toprule
                        & \multicolumn{5}{c}{End-Diastole (ED) phase} & \multicolumn{5}{c}{End-Systole (ES) phase} \\
 Label & DSC    & Jaccard & HD      & MSD    & HD95   & DSC    & Jaccard & HD      & MSD    & HD95   \\
\midrule
 LV    & 0.968  & 0.938   & 4.082  & 0.871  & 2.977  & 0.948  & 0.904   & 4.432  & 1.076  & 3.246  \\
 MYO   & 0.878  & 0.786   & 6.504  & 0.662  & 2.151  & 0.891  & 0.809   & 5.342  & 0.837  & 2.809  \\
 RV    & 0.934  & 0.878   & 6.055  & 1.262  & 4.075  & 0.899  & 0.822   & 6.108  & 1.457  & 4.254  \\
\bottomrule
\end{tabular}
\end{table}
\begin{table}[!h]
\caption{Comparison of Dice scores of nnUnet and Other methods. ED - End Diastole, ES - End Systole, ShA- Short Axis, LoA - Long Axis, LV - Left Ventricle, MYO - Myocardium, RV - Right Ventricle, LA - Left Atrium. }
\label{table:main_nnunet_comparison}
\centering
\begin{tabular}{lccccc}
\hline
\multirow{2}{*}{Dataset} &  \multirow{2}{*}{Sub task} &

\multirow{2}{*}{Label}
 & \multirow{2}{*}{nnU-Net} &  \multirow{2}{*}{\parbox[t]{1.2cm}{Other\\ Methods}} &
 \multirow{2}{*}{\parbox[t]{2cm}{Abs\\ Difference (\%) }} \\
 & & & & &\\
\midrule

% \multirow{2}{0pt}{Utah} & - & LA & \textbf{0.921}&0.919 &0.2 \\
%                         & - & RA & 0.915& 0.921&0.6 \\
%                         \midrule

\multirow{6}{*}{ACDC} &  ED &\multirow{2}{*}{LV}  & 0.965& 0.968& 0.3\\
                        &  ES &  & 0.927& 0.938& 1.1\\ \cmidrule(lr){2-6}
                        &  ED &\multirow{2}{*}{MYO}  & 0.898& 0.906& 0.8\\
                        &  ES &  & 0.915& 0.923& 0.8\\ \cmidrule(lr){2-6}
                        &  ED &\multirow{2}{*}{RV}  & 0.944& 0.955& 1.1\\
                        &  ES&  & 0.892& 0.904& 1.2\\ \midrule

\multirow{3}{0pt}{LAScarQS} &  Task-1& LA Scar& 0.449& 0.660&21.1\\
  &  Task-1 &LA Cavity& \textbf{0.939}& 0.938&0.1\\
  &  Task-2& LA Cavity& \textbf{0.938}& 0.929& 0.9\\ \midrule
\multirow{6}{0pt}{MnM1} & ED& \multirow{2}{*}{LV}& 0.937& 0.940&0.3\\
 & ES& & \textbf{0.909}& 0.890&1.9\\ \cmidrule(lr){2-6}
 & ED& \multirow{2}{*}{MYO} & 0.826& 0.834&0.8\\
 & ES&  & 0.864& 0.870&0.6\\ \cmidrule(lr){2-6}
 & ED & \multirow{2}{*}{RV}  & \textbf{0.913}& 0.910&0.3\\
 & ES&   & \textbf{0.893}& 0.860&3.3\\ \midrule
 
\multirow{4}{0pt}{MnM2} &  ShA ED&\multirow{4}{*}{RV}& 0.937& 0.940&0.3\\
&  ShA ES& & \textbf{0.938}& 0.914&2.4\\
 &  LoA ES& & 0.934& 0.935&0.1\\
 &  LoA ED& & 0.900& 0.905&0.5\\ \midrule

 LASC &  -&LA Cavity& 0.934& 0.935& 0.1\\
 \bottomrule
\end{tabular}
\end{table}

Deviating from the MnM challenge, we analyze the performance of the MnM2 challenge in four different conditions: Short Axis (ShA) ED phase, ShA ES phase (Table~\ref{tab:mnm2_SA_merged}), Long Axis (LoA) ED phase and LoA ES phase (Table~\ref{tab:mnm2_LA_merged}). In both phases in ShA, the ensemble method demonstrates superior performance (DSC-0.937). For LoA segmentation, images have the shape of $H \times W \times 1$, indicating only one layer in the Z-axis, making the 3D full-resolution method particularly effective, and thus only 3D full-resolution results are reported. The challenge organizers report only the values of RV segmentation (Table~\ref{tab:MnM2_compare}). In this case, nnU-Net outperforms ShA ES segmentation by 2.4\% compared to other models in DSC (0.914 vs 0.938). However, in other cases (ShA ED (0.940 vs 0.937), LoA ES (0.905 vs 0.900), and LoA ED (0.935 vs 0.934)), other models surpass the nnU-Net, but the margin is less than 1\%. In the ShA view, both ED and ES have similar performance in DSC and Jaccard while in LoA, overall LV and RV have better results in the ED phase compared to the ES phase and MYO has better performance in the ES phase compared to the ED phase.

In summary, ensemble models demonstrate strong performance across all datasets. Surprisingly, in some cases, the 2D models outperform the 3D models and even the ensemble models. The most significant difference where other models surpass nnU-Net occurs in the LAScarQs Task 1 scar segmentation. A summary of the comparison between the highest Dice value obtained from nnU-Net, the highest Dice value from other methods, and the absolute difference (\%) is shown in Table~\ref{table:main_nnunet_comparison}.

\section{Discussion}

In this section, we discuss and analyze our findings in detail.

\subsection{Lower performance in LAScarQS scar segmentation}

In analyzing the performance of nnU-Net for scar segmentation in the LAScarQS Task 1, it is evident that the nnU-Net underperforms relative to other models. Several factors contribute to this discrepancy. Firstly, the primary challenge lies in the nature of the target region. Scar tissues occupy only a small fraction of the LA compared to the LA cavity (As shown in Fig.~\ref{fig:sup_pix_dist} in supplementary nearly 0.7\% occupies the cavity, and less than 0.1\% occupies the scar). This significant imbalance in the spatial distribution makes it difficult for the model to accurately distinguish and segment the scar regions. The nnU-Net's architecture, while robust for larger and more continuous regions, struggles with the precision required for such minute and sparse areas.

Secondly, the characteristics of the data further complicate the task. Unlike the LA cavity, which presents as a more continuous and homogenous region, scar tissues are often irregular and dispersed. This non-continuous nature of scar data poses a substantial challenge for segmentation models, particularly those like nnU-Net, which rely heavily on spatial continuity and context provided by larger regions.

Additionally, most state-of-the-art methods for scar segmentation adopt a two-stage network approach. These approaches typically involve an initial stage that performs coarse segmentation, identifying potential regions of interest (ROIs), followed by a refinement stage that focuses on enhancing the segmentation accuracy within these regions. This two-step process allows for more focused learning and better handling of small and irregular regions, leading to superior performance in scar segmentation tasks. In contrast, the nnU-Net framework primarily utilizes a single-stage approach. While this method is advantageous for its simplicity and reduced computational requirements, it may not provide the necessary granularity and focus required for effectively segmenting small and irregular structures like scar tissues. The lack of an initial coarse segmentation stage means that nnU-Net must rely solely on its inherent ability to capture and distinguish fine details within a single pass, which is inherently more challenging for such complex tasks.

Moreover, the non-continuous property of the scar tissue can contribute to higher HD values. The HD metric is particularly sensitive to outliers and disjoint regions, which are characteristic of scar tissue. As a result, even small segmentation errors can lead to disproportionately high HD values, further reflecting the difficulty in accurately segmenting these regions.

% Lastly, the standard data augmentation and preprocessing techniques employed by nnU-Net, while effective for general segmentation tasks, might not be sufficiently tailored to the unique challenges presented by scar tissue segmentation. Employing more specialized augmentation techniques that better simulate the variability and appearance of scar tissues could potentially enhance the model's performance.

\subsection{Ensemble Results}

When comparing nnU-Net ensemble models to individual 3D and 2D nnU-Net variants, it is essential to understand that while ensemble methods have the potential to enhance model performance, this improvement is not always guaranteed. For an ensemble to significantly outperform a single model, the base classifiers must exhibit diversity. This means they need to make different errors, thereby complementing each other's weaknesses. However, when the signal in the data is dominated by a few strong predictors, most models, including those within an ensemble, will likely capture and model this dominant information similarly. This can result in highly correlated predictions across the ensemble members, thereby reducing the potential benefits of combining them. In other words, if the nnU-Net ensemble models demonstrate lower performance compared to individual 3D or 2D nnU-Net variants, a lack of diversity among the ensemble members could be a contributing factor. When ensemble models are not sufficiently diverse, they may fail to provide the expected performance boost, leading to a situation where the ensemble's performance is merely on par with or even inferior to the best individual model.

\subsection{Higher performance in 2D model compared to 3D model}

In our analysis of the ACDC, MnM, and MnM2 datasets, we observe a trend where 2D nnU-Net implementations demonstrated superior performance, as measured by Dice scores, compared to their 3D counterparts. While no significant differences (all statistical tests were conducted using the Wilcoxen test) were observed between 2D and 3D nnU-Net implementations for LV and RV segmentation, the MYO shows statistically significant results in those configurations. Specifically, differences were observed in ACDC (ED and ES phases), the MnM (ED phase) and the MnM2(ES phase). There was no statistical significance in the MnM-ES phase and MnM2-ED phase even though the 2D model shows a higher dice score. These results suggest that the myocardium poses unique challenges and opportunities for segmentation, distinguishing it from the LV and RV. 

One of the main reasons would be the myocardium's structure and imaging characteristics, which make its segmentation inherently different from that of the LV and RV. Unlike the LV and RV, which are more clearly defined by their lumen boundaries, the myocardium often has less distinct edges due to variability in contrast and partial volume effects. This nuanced texture and boundary information may favour the use of simpler 2D architectures, which can focus on in-plane details without being overwhelmed by 3D spatial relationships that may introduce noise.

Also, MRI data typically exhibit high in-plane resolution but lower through-plane resolution. For the myocardium, where subtle edge and texture features are critical for accurate segmentation, the higher in-plane resolution plays a key role. 2D models excel in leveraging these high-resolution details, whereas 3D models may struggle to integrate lower-resolution through-plane information effectively.

3D models, by their nature, require more parameters and training data to generalize effectively. This requirement becomes particularly challenging for the myocardium, where segmentation demands a fine-grained understanding of complex features. In cases where the dataset size is limited, 2D models may achieve better generalization due to their lower parameter count and simpler optimization landscape.

Furthermore, the nature of the segmentation task itself may favour 2D approaches. If the key features for accurate segmentation are predominantly visible within individual slices, the additional complexity introduced by 3D models in capturing inter-slice relationships may not provide significant benefits. In fact, this added complexity could potentially introduce noise or irrelevant information into the learning process, leading to suboptimal performance.

% #### 4. Ensemble Model Limitations
% For the MnM dataset, differences were observed between 2D models and ensembles for the myocardium. Ensembles typically aim to combine the strengths of multiple models, but in this case, the added complexity may dilute the specific strengths of 2D models that are well-suited to capturing the myocardium's in-plane features. This could explain why ensembles did not outperform 2D models for MYO segmentation.

% #### 5. Task-Specific Challenges
% The segmentation of the myocardium often relies on subtle, localized features that may not benefit as much from the volumetric context provided by 3D models. Instead, this additional context may introduce irrelevant inter-slice information, reducing the model's focus on the crucial in-plane features that define the myocardium's boundaries.

\subsection{Effect of the configurations of nnU-Net}
When utilizing nnU-Net, the selection of loss functions, optimizers, batch sizes, and patch sizes is tailored to the specific characteristics of the dataset. In our case, all the nnU-Nets employ a combination of Dice loss and cross-entropy loss (DiceCE loss) as its default loss function. However, in scenarios with class imbalance, alternative loss functions such as DiceHD loss (combining Dice loss with Hausdorff Distance loss) and DiceFocal loss (combining Dice loss with focal loss) have demonstrated superior performance~\citep{ma2021loss}. Therefore, incorporating these loss functions into nnU-Net could potentially enhance segmentation results.

Furthermore, nnU-Net traditionally utilises the SGD optimizer. Nonetheless, recent studies have shown that the Adam optimizer can achieve comparable, if not superior, outcomes in segmentation tasks~\cite {rajinikanth2022skin}. Consequently, integrating the Adam optimizer into nnU-Net's framework may lead to improved performance in certain cases.

\subsection{Performance Differences in End Systole and End Diastole Phases}

The results indicate that for the ACDC and MnM datasets, the LV and RV exhibit better segmentation performance during the End-Diastole (ED) phase, particularly in Dice and Jaccard scores. In contrast, the Myocardium (MYO) achieves better segmentation performance during the End-Systole (ES) phase.

In the ED phase, both LV and RV are at their maximum size due to blood filling, which results in more distinct boundaries and reduced segmentation errors. Larger structures with smoother curvature and higher contrast against surrounding tissues are generally easier to segment. Additionally, the heart experiences reduced motion artifacts during ED due to slower movement, further improving segmentation accuracy. Conversely, during the ES phase, the ventricles contract and become smaller, leading to less distinct boundaries and increased structural deformation. These factors introduce variability that complicates the segmentation task for the LV and RV.

However, the MYO achieves better segmentation in the ES phase, primarily due to its increased thickness during contraction. The enhanced visibility and prominence of the thickened myocardium, along with its well-defined boundaries, aid in achieving higher segmentation accuracy. In contrast, during the ED phase, the thinner myocardium has less distinct boundaries, making it more challenging to segment accurately.

For the MnM2 dataset, the segmentation performance in ED and ES phases is comparable, particularly in Dice and Jaccard scores. This consistency can be attributed to the imaging modality, which captures the heart in the ShA view. In this view, the heart's structures, including the LV, RV, and MYO, maintain relatively similar cross-sectional shapes across both phases. Volume changes during the cardiac cycle predominantly occur along the long axis (apex-base direction), which is less apparent in the Short Axis view. This uniformity reduces structural distortions and phase-dependent variability. 

\subsection{Limitations of the Study}

This study primarily focused on cardiac-related datasets and utilized a single imaging modality, namely cardiac MRI. While this approach allowed for a detailed evaluation of nnU-Net’s performance in a specific domain, it limits the generalizability of the findings. Future research should extend this work to encompass other anatomical regions, such as the brain and abdomen, which are commonly studied in medical imaging. Additionally, incorporating other imaging modalities, including computed tomography (CT), X-ray, and ultrasound, would provide a broader validation of nnU-Net's robustness and highlight its applicability across diverse clinical scenarios. Exploring these modalities could uncover unique challenges or advantages specific to nnU-Net’s architecture when applied to different data types.

Another limitation is the study’s focus on the baseline nnU-Net architecture without exploring existing variations or improvements made by other researchers~\citep{isensee2024nnu,jahromi2024nnu}. Evaluating these architectural enhancements could provide valuable insights into their impact on segmentation performance and computational efficiency, thereby enriching the comparative analysis.

Moreover, this study exclusively relied on publicly available datasets that are widely used within the research community. While this ensures reproducibility and relevance, the findings may not fully represent the challenges encountered in real-world clinical data, which can vary significantly in terms of quality, resolution, and heterogeneity. Future studies could benefit from incorporating private datasets or datasets with more diverse patient populations and imaging conditions to better understand nnU-Net’s performance in clinical practice.

Finally, our study did not focus on ventricular scars. Ventricular scar segmentation is crucial for many clinical applications, such as planning cardiac ablation procedures and assessing myocardial viability. Future work could explore this aspect in greater depth, utilizing dedicated datasets and task-specific modifications to evaluate nnU-Net’s capability to handle these clinically significant tasks.

\section{Conclusion}

In this study, we evaluated the performance of five cardiac MRI segmentation datasets using various adaptations of nnU-Net. Through over 130 training cycles, we conducted an extensive performance analysis of these models. Our comparative study with existing methods demonstrated that nnU-Net not only performs competitively but often surpasses state-of-the-art techniques, even outperforming the latest methods on certain datasets.

The findings highlight the robustness and adaptability of nnU-Net for cardiac MRI segmentation tasks. Its consistent performance across diverse datasets underscores its potential as a reliable tool for clinical applications. However, this study raises an important question: when is it necessary to develop task-specific models for particular cardiac segmentation challenges?

The answer lies in the specific requirements and complexities of individual tasks. While nnU-Net provides a strong baseline, certain scenarios may demand customized solutions. For example, while nnU-Net excels in segmenting larger anatomical structures such as the LA cavity, LV, and RV, its performance in segmenting more intricate regions, such as the MYO and scars, is relatively lower. In such cases, developing specialized models can lead to improved outcomes. Furthermore, integrating data from multiple imaging modalities (e.g., MRI and CT) may require tailored approaches to effectively interpret and merge the information.

In conclusion, while nnU-Net offers a robust and versatile foundation for cardiac MRI segmentation, the development of specialized models tailored to specific clinical challenges remains essential. This study demonstrates that while general-purpose frameworks like nnU-Net provide significant advantages, there is a critical need for continued innovation and customization to address the unique complexities of various medical imaging tasks. Future research should prioritize the exploration and development of these specialized approaches to fully leverage the potential of deep learning in medical imaging.

\section{List of abbreviations}

 \begin{itemize}
 \item CVD - Cardiovascular diseases
     \item AF - Atrail Fibrilation
     \item MRI - Magnetic Resonance Imaging
     \item LA - Left Aria
     \item MYO - Myocardium
     \item RA - Right Atria
     \item LV - Left Ventricle
     \item RV - Right Ventricle
     \item ES - End Systole
     \item ED - End Diastole
     \item ShA - Short Axis
     \item LoA - Long Axis
     \item DSC - Dice Score
     \item HD - Hausdorff Distance
     \item MSD - Mean Surface Distance
 \end{itemize}

\section{Declarations}

\subsection{Ethics approval and consent to participate}

Not applicable

\subsection{Consent for publication}

Not applicable

\subsection{Availability of data and materials}
The datasets analysed during the current study are publically available in the following repositories except for the LAScarQs 2022 dataset \citep{zhuang2023left}, which was obtained from the corresponding authors of the challenge.
\\
LASC \citep{xiong2021global}:- https://www.cardiacatlas.org/atriaseg2018-challenge/atria-seg-data/ \\
ACDC \citep{bernard2018deep}:- https://www.creatis.insa-lyon.fr/Challenge/acdc/databases.html \\
MnM1 \citep{campello2021multi}:- https://www.ub.edu/mnms/ \\
MnM2 \citep{martin2023deep}:- https://www.ub.edu/mnms-2/ \\

\subsection{Competing Interests}
Not applicable

\subsection{Funding}
Not applicable

\subsection{Author contributions}
MG is the main author of the paper, who conducted the experiments and wrote the manuscript. FX contributed by conducting experiments and preparing tables. JZ served as the research advisor and prepared figures. All authors review the article. 

\subsection{Acknowledgements}
Not applicable

\bibliography{sn-bibliography}
% Supplementary Section
\clearpage % Start on a new page
\section*{Additional file 1}
\setcounter{section}{0} % Reset section counter
\renewcommand{\thesection}{A\arabic{section}} % Add "S" to section numbers
\renewcommand{\thefigure}{A\arabic{figure}} % Add "S" to figure numbers
\renewcommand{\thetable}{A\arabic{table}} % Add "S" to table numbers
\counterwithin{equation}{section} % Equations in the supplementary start with "S"
\setcounter{figure}{0} % Reset figure counter
\setcounter{table}{0} % Reset table counter

\section{Evaluation Metrics}
\label{sup:eval_metrics}
Here we explained the evaluation metrics in detail.

\subsubsection{Dice Score}
The Dice Similarity Coefficient (DSC) is a measure of overlap between the predicted segmentation and the ground truth, calculated as twice the area of overlap divided by the total number of pixels in both the predicted and ground truth masks. A higher DSC indicates better performance, signifying a greater degree of similarity between the predicted and actual segmentation.

\begin{equation}
DSC = \frac{2 \cdot |P \cap Q|}{|P| + |Q|}
\end{equation}
where P and Q are the ground truth and predicted masks.

\subsubsection{Jaccard Index}
The Jaccard Index, also known as the Intersection over Union (IoU), quantifies the similarity between the predicted and ground truth segmentation. It is defined as the area of overlap divided by the area of the union of the predicted and ground truth masks. Similar to the DSC, a higher Jaccard Index denotes better segmentation performance.

\begin{equation}
Jaccard = \frac{|P \cap Q|}{|P \cup Q|}
\end{equation}
where P and Q are the ground truth and predicted masks.

\subsubsection{Hausdorff Distance }

The Hausdorff Distance (HD) measures the maximum distance from a point in the predicted segmentation to the nearest point in the ground truth segmentation, thus indicating the worst-case boundary discrepancy. Lower HD values indicate more accurate boundary delineation.

\begin{equation}
HD(P, Q) = \max(h(P, Q), h(Q, P))
\end{equation}
where \( h(P, Q) \) is the oriented Hausdorff distance from \( P \) to \( Q \):

\begin{equation}
h(P, Q) = \max_{p_i \in P} \min_{q_j \in Q} \rho(p_i, q_j)
\end{equation}

and \( \rho(p_i, q_j) \) is the Euclidean distance between points \( p_i \) and \( q_j \).

\subsubsection{Mean Surface Distance}

The Mean Surface Distance (MSD) calculates the average distance between points on the surface of the predicted segmentation and the nearest points on the surface of the ground truth segmentation. Lower MSD values suggest closer average alignment between the predicted and actual boundaries.

\begin{equation}
MSD(P, Q) = \frac{1}{|P|} \sum_{p_i \in P} \min_{q_j \in Q} \rho(p_i, q_j)
\end{equation}

\subsubsection{95th percentile Hausdorff Distance}

The 95th percentile Hausdorff Distance (HD95) is similar to the HD but focuses on the 95th percentile of the distances between the predicted and ground truth surfaces, thereby mitigating the impact of outliers. A lower HD95 value indicates more consistent boundary accuracy, discounting extreme deviations.

Together, these metrics provide a robust framework for evaluating segmentation performance, with higher DSC and Jaccard Index values and lower HD, MSD, and HD95 values indicating superior model performance.

\section{Datasets}
\label{sup:dataset}

A summary of the datasets is shown in Table \ref{tab:sup_summary_cardiac_datasets}.
\begin{table}[h]
 \caption{Summary of Cardiac MRI Datasets}
    \label{tab:sup_summary_cardiac_datasets}
    \centering
    \begin{tabular}{ lllcc }
        \toprule
        Dataset &  Task& Labels & Training & Testing \\
        \midrule

        \multirow{2}{*}{LAScarQS} 
        & Task 1 & LA cavity, scars & 50& 10 \\
        & Task 2 & LA cavity & 130 & 20 \\
       \midrule
        \multirow{1}{*}{LASC} 
        &  - &LA cavity & 100 & 54 \\
        \midrule
        \multirow{2}{*}{ACDC}&  End-Diastole & \multirow{2}{*}{LV, MYO, RV} & 100 & 50 \\
 & End-Systole& & 100&50\\
       \midrule
        \multirow{2}{*}{MnM-1}&  End-Diastole
& \multirow{2}{*}{LV, MYO, RV} & 150& 136 \\
 & End-Systole& & 150&136 \\
      \midrule
        \multirow{4}{*}{MnM-2}&  Short Axis, End-Diastole& \multirow{4}{*}{LV, MYO, RV} & 200 & 160 \\
 & Short Axis, End-Systole& & 200 &160 \\
 & Long Axis, End-Diastole& & 200 &160 \\
 & Long Axis, End-Systole& & 200 &160 \\
 \bottomrule
    \end{tabular}
\end{table}

\section{Additional Results}

\subsection{Validation set and test set results}

In Tables~\ref{tab:sup_validaiton} and~\ref{tab:sup_test}, we present the performance results for the validation and test sets, reported as mean $\pm$ standard deviation. For the validation set, the results are averaged across the five folds, whereas for the test set, the averages are computed over all test images.

\begin{center}
\begin{longtable}{p{1.5cm}|p{1.1cm}|p{1.3cm}|p{1.3cm}|p{2.4cm}|p{2.4cm}}
\caption{Validation set performance comparison across LAScarQS, LASC, ACDC, MnM, and MnM2 datasets, reported as mean $\pm$ std. Abbreviations: Config. Configuration, LAC - Left Atrial Cavity, RV - Right Ventricle, MYO - Myocardium, LV - Left Ventricle, ED - End Diastole, ES - End Systole, full - full resolution, low - low resolution, cas. - cascade.} \label{tab:sup_validaiton} \\
\toprule
\textbf{Dataset} & \textbf{Task} & \textbf{Config.} & \textbf{Label} & \textbf{Dice } & \textbf{HD95 } \\
\midrule
\endfirsthead

\caption{(continued)} \\
\toprule
\textbf{Dataset} & \textbf{Task} & \textbf{Config.} & \textbf{Label} & \textbf{Dice } & \textbf{HD95 } \\
\midrule
\endhead

\bottomrule
\endfoot

\endlastfoot

\multirow{4}{*}{LAScarQS} & Task2 & 2D & LAC & $0.921 \pm 0.0080$ & $3.866 \pm 0.4962$ \\
          &       & 3D full. & LAC & $0.925 \pm 0.0071$ & $3.707 \pm 0.3914$ \\
          &       & 3D low. & LAC & $0.927 \pm 0.0057$ & $3.454 \pm 0.2390$ \\
          &       & 3D cas. & LAC & $0.926 \pm 0.0054$ & $3.621 \pm 0.3089$ \\ \cmidrule{2-6}
          & Task1 & 2D & LAC & $0.900 \pm 0.0244$ & $4.427 \pm 0.6797$ \\
          &       &     & Scar & $0.493 \pm 0.0524$ & $7.942 \pm 1.6955$ \\ \cmidrule{3-6}
          &       & 3D full. & LAC & $0.923 \pm 0.0131$ & $3.334 \pm 0.3076$ \\
          &       &     & Scar & $0.501 \pm 0.0452$ & $7.898 \pm 1.7594$ \\ \cmidrule{3-6}
          &       & 3D low. & LAC & $0.923 \pm 0.0116$ & $3.346 \pm 0.4136$ \\
          &       &     & Scar & $0.474 \pm 0.0496$ & $8.554 \pm 1.6419$ \\ \cmidrule{3-6}
          &       & 3D cas. & LAC & $0.923 \pm 0.0093$ & $3.612 \pm 0.4662$ \\
          &       &     & Scar & $0.472 \pm 0.1110$ & $8.662 \pm 3.0562$ \\
\hline
LASC      & -     & 2D & LAC & $0.909 \pm 0.0199$ & $4.830 \pm 0.8469$ \\
          &       & 3D full. & LAC & $0.927 \pm 0.0036$ & $4.079 \pm 0.2672$ \\
          &       & 3D low. & LAC & $0.926 \pm 0.0043$ & $3.984 \pm 0.3186$ \\
          &       & 3D cas. & LAC & $0.927 \pm 0.0041$ & $4.032 \pm 0.3590$ \\
\hline
ACDC      & ED    & 2D & RV & $0.935 \pm 0.0053$ & $3.461 \pm 0.6634$ \\
          &       &     & MYO & $0.894 \pm 0.0092$ & $1.693 \pm 0.2132$ \\
          &       &     & LV & $0.964 \pm 0.0029$ & $2.454 \pm 0.7448$ \\ \cmidrule{3-6}
          &       & 3D full. & RV & $0.935 \pm 0.0053$ & $3.461 \pm 0.6634$ \\
          &       &     & MYO & $0.894 \pm 0.0092$ & $1.693 \pm 0.2132$ \\
          &       &     & LV & $0.964 \pm 0.0029$ & $2.454 \pm 0.7448$ \\ \cmidrule{2-6}
          & ES    & 2D & RV & $0.871 \pm 0.0167$ & $5.110 \pm 1.0900$ \\
          &       &     & MYO & $0.904 \pm 0.0048$ & $2.244 \pm 0.3354$ \\
          &       &     & LV & $0.923 \pm 0.0111$ & $4.487 \pm 1.1083$ \\ \cmidrule{3-6}
          &       & 3D full. & RV & $0.852 \pm 0.0113$ & $6.053 \pm 0.4896$ \\
          &       &     & MYO & $0.901 \pm 0.0080$ & $2.348 \pm 0.4481$ \\
          &       &     & LV & $0.924 \pm 0.0139$ & $3.137 \pm 0.5750$ \\
\hline
MnM       & ED    & 2D & LV & $0.947 \pm 0.0194$ & $3.658 \pm 1.2410$ \\
          &       &     & MYO & $0.861 \pm 0.0267$ & $3.666 \pm 1.6247$ \\
          &       &     & RV & $0.919 \pm 0.0110$ & $4.304 \pm 1.0132$ \\ \cmidrule{3-6}
          &       & 3D full. & LV & $0.950 \pm 0.0041$ & $3.285 \pm 0.3301$ \\
          &       &     & MYO & $0.861 \pm 0.0105$ & $3.041 \pm 0.5260$ \\
          &       &     & RV & $0.908 \pm 0.0114$ & $5.116 \pm 0.9079$ \\\cmidrule{2-6}
          & ES    & 2D & LV & $0.903 \pm 0.0108$ & $4.442 \pm 0.8087$ \\
          &       &     & MYO & $0.872 \pm 0.0165$ & $3.290 \pm 0.9736$ \\
          &       &     & RV & $0.863 \pm 0.0208$ & $4.844 \pm 0.5368$ \\ \cmidrule{3-6}
          &       & 3D full. & LV & $0.902 \pm 0.0078$ & $4.055 \pm 0.3486$ \\
          &       &     & MYO & $0.867 \pm 0.0098$ & $3.323 \pm 0.4579$ \\
          &       &     & RV & $0.851 \pm 0.0170$ & $5.629 \pm 1.0387$ \\
\hline
MnM2      & LA ED & 3D full. & LV & $0.969 \pm 0.0032$ & $2.597 \pm 0.2565$ \\
          &       &     & MYO & $0.872 \pm 0.0101$ & $1.937 \pm 0.2253$ \\
          &       &     & RV & $0.936 \pm 0.0065$ & $4.355 \pm 1.0182$ \\ \cmidrule{2-6}
          & LA ES & 3D full. & LV & $0.956 \pm 0.0066$ & $2.780 \pm 0.3659$ \\
          &       &     & MYO & $0.895 \pm 0.0122$ & $2.190 \pm 0.3762$ \\
          &       &     & RV & $0.909 \pm 0.0161$ & $4.207 \pm 1.2462$ \\ \cmidrule{2-6}
          & SA ED & 2D & LV & $0.958 \pm 0.0032$ & $3.672 \pm 0.6647$ \\
          &       &     & MYO & $0.872 \pm 0.0056$ & $2.761 \pm 0.3829$ \\
          &       &     & RV & $0.940 \pm 0.0086$ & $3.961 \pm 1.1052$ \\ \cmidrule{3-6}
          &       & 3D full. & LV & $0.961 \pm 0.0029$ & $3.163 \pm 0.3944$ \\
          &       &     & MYO & $0.872 \pm 0.0076$ & $2.427 \pm 0.3216$ \\
          &       &     & RV & $0.935 \pm 0.0081$ & $4.780 \pm 1.1574$ \\ \cmidrule{2-6}
          & SA ES & 2D & LV & $0.958 \pm 0.0025$ & $3.804 \pm 0.6177$ \\
          &       &     & MYO & $0.870 \pm 0.0081$ & $2.913 \pm 0.2344$ \\
          &       &     & RV & $0.943 \pm 0.0099$ & $3.426 \pm 0.9041$ \\ \cmidrule{3-6}
          &       & 3D full. & LV & $0.957 \pm 0.0025$ & $3.563 \pm 0.2242$ \\
          &       &     & MYO & $0.871 \pm 0.0100$ & $2.580 \pm 0.3922$ \\
          &       &     & RV & $0.939 \pm 0.0084$ & $4.038 \pm 1.0009$ \\
\hline
\end{longtable}
\end{center}

\begin{center}
\begin{longtable}{p{1.5cm}|p{1.1cm}|p{1.3cm}|p{1.3cm}|p{2.4cm}|p{2.4cm}}
\caption{Test set performance comparison across LAScarQS, LASC, ACDC, MnM, and MnM2 datasets, reported as mean $\pm$ std. Abbreviations: Config. Configuration, LAC - Left Atrial Cavity, RV - Right Ventricle, MYO - Myocardium, LV - Left Ventricle, ED - End Diastole, ES - End Systole, full - full resolution, low - low resolution, cas. - cascade, Ens.- Ensemble.} \label{tab:sup_test} \\
\toprule
\textbf{Dataset} & \textbf{Task} & \textbf{Config.} & \textbf{Label} & \textbf{Dice } & \textbf{HD95 } \\
\midrule
\endfirsthead

\caption{(continued)} \\
\toprule
\textbf{Dataset} & \textbf{Task} & \textbf{Config.} & \textbf{Label} & \textbf{Dice } & \textbf{HD95 } \\
\midrule
\endhead

\bottomrule
\endfoot

\endlastfoot

LAScarQS & Task2 & 2D & LAC & $0.930 \pm 0.0168$ & $3.017 \pm 0.9052$ \\
          &       & 3D  full. & LAC & $0.937 \pm 0.0149$ & $2.880 \pm 0.9661$ \\
          &       & 3D low. & LAC & $0.935 \pm 0.0164$ & $3.069 \pm 1.3614$ \\
          &       & 3D cas. & LAC & $0.937 \pm 0.0161$ & $2.745 \pm 0.8492$ \\
          &       & Ens. & LAC & $0.938 \pm 0.0151$ & $2.737 \pm 0.8157$ \\ \cmidrule{2-6}
          & Task1 & 2D & LAC & $0.926 \pm 0.0151$ & $3.402 \pm 1.1058$ \\
          &       &     & Scar & $0.438 \pm 0.0755$ & $13.036 \pm 5.7312$ \\ \cmidrule{3-6}
          &       & 3D  full. & LAC & $0.939 \pm 0.0099$ & $3.088 \pm 1.0105$ \\
          &       &     & Scar & $0.443 \pm 0.0911$ & $12.620 \pm 3.6403$ \\ \cmidrule{3-6}
          &       & 3D low. & LAC & $0.937 \pm 0.0126$ & $3.254 \pm 1.2967$ \\
          &       &     & Scar & $0.411 \pm 0.0841$ & $13.425 \pm 3.5387$ \\ \cmidrule{3-6}
          &       & 3D cas. & LAC & $0.939 \pm 0.0096$ & $3.138 \pm 1.1646$ \\
          &       &     & Scar & $0.449 \pm 0.0852$ & $12.554 \pm 3.6200$ \\ \cmidrule{3-6}
          &       & Ens. & LAC & $0.939 \pm 0.0105$ & $3.041 \pm 1.0385$ \\
          &       &     & Scar & $0.439 \pm 0.0878$ & $12.850 \pm 3.6996$ \\
\hline
LASC      & -     & 2D & LAC & $0.926 \pm 0.0227$ & $3.930 \pm 2.1143$ \\
          &       & 3D  full. & LAC & $0.933 \pm 0.0202$ & $3.681 \pm 2.0243$ \\
          &       & 3D low. & LAC & $0.931 \pm 0.0214$ & $3.727 \pm 2.0254$ \\
          &       & 3D cas. & LAC & $0.933 \pm 0.0210$ & $3.756 \pm 2.0699$ \\
          &       & Ens. & LAC & $0.934 \pm 0.0204$ & $3.628 \pm 2.0353$ \\
\hline
ACDC      & ED    & 2D & RV & $0.942 \pm 0.0361$ & $3.152 \pm 2.7885$ \\
          &       &     & MYO & $0.897 \pm 0.0197$ & $1.583 \pm 0.2452$ \\
          &       &     & LV & $0.965 \pm 0.0170$ & $2.350 \pm 2.2734$ \\ \cmidrule{3-6}
          &       & 3D  full. & RV & $0.934 \pm 0.0469$ & $3.861 \pm 4.0508$ \\
          &       &     & MYO & $0.889 \pm 0.0218$ & $2.135 \pm 1.9664$ \\
          &       &     & LV & $0.959 \pm 0.0258$ & $2.720 \pm 2.6258$ \\ \cmidrule{3-6}
          &       & Ens. & RV & $0.944 \pm 0.0367$ & $3.110 \pm 2.8565$ \\
          &       &     & MYO & $0.898 \pm 0.0194$ & $1.818 \pm 1.4122$ \\
          &       &     & LV & $0.963 \pm 0.0212$ & $2.474 \pm 2.3381$ \\ \cmidrule{2-6}
          & ES    & 2D & RV & $0.885 \pm 0.0578$ & $4.318 \pm 3.7865$ \\
          &       &     & MYO & $0.913 \pm 0.0245$ & $1.975 \pm 1.4111$ \\
          &       &     & LV & $0.927 \pm 0.0496$ & $2.713 \pm 2.6216$ \\ \cmidrule{3-6}
          &       & 3D  full. & RV & $0.882 \pm 0.0591$ & $5.245 \pm 3.7910$ \\
          &       &     & MYO & $0.906 \pm 0.0239$ & $2.551 \pm 2.4158$ \\
          &       &     & LV & $0.901 \pm 0.0884$ & $4.968 \pm 6.0382$ \\ \cmidrule{3-6}
          &       & Ens. & RV & $0.892 \pm 0.0555$ & $4.208 \pm 3.8025$ \\
          &       &     & MYO & $0.915 \pm 0.0211$ & $2.193 \pm 1.9121$ \\
          &       &     & LV & $0.922 \pm 0.0601$ & $3.477 \pm 3.3692$ \\
\hline
MnM       & ED    & 2D & LV & $0.936 \pm 0.0752$ & $3.871 \pm 5.5988$ \\
          &       &     & MYO & $0.824 \pm 0.0537$ & $3.676 \pm 5.1438$ \\
          &       &     & RV & $0.909 \pm 0.0823$ & $4.578 \pm 5.0231$ \\ \cmidrule{3-6}
          &       & 3D  full. & LV & $0.933 \pm 0.0678$ & $4.261 \pm 3.4927$ \\
          &       &     & MYO & $0.819 \pm 0.0451$ & $3.484 \pm 3.1505$ \\
          &       &     & RV & $0.908 \pm 0.0906$ & $4.501 \pm 5.3539$ \\ \cmidrule{3-6}
          &       & Ens. & LV & $0.937 \pm 0.0530$ & $3.761 \pm 4.1780$ \\
          &       &     & MYO & $0.826 \pm 0.0645$ & $3.138 \pm 3.3104$ \\
          &       &     & RV & $0.913 \pm 0.0865$ & $4.208 \pm 4.5919$ \\ \cmidrule{2-6}
          & ES    & 2D & LV & $0.888 \pm 0.0380$ & $8.513 \pm 3.6098$ \\
          &       &     & MYO & $0.800 \pm 0.0648$ & $7.304 \pm 4.6097$ \\
          &       &     & RV & $0.893 \pm 0.0526$ & $6.411 \pm 3.8556$ \\ \cmidrule{3-6}
          &       & 3D  full. & LV & $0.909 \pm 0.0429$ & $4.507 \pm 4.0052$ \\
          &       &     & MYO & $0.841 \pm 0.0651$ & $3.952 \pm 3.1857$ \\
          &       &     & RV & $0.871 \pm 0.0515$ & $5.673 \pm 3.7000$ \\ \cmidrule{3-6}
          &       & Ens. & LV & $0.888 \pm 0.0405$ & $4.432 \pm 3.2381$ \\
          &       &     & MYO & $0.864 \pm 0.0486$ & $3.542 \pm 3.4477$ \\
          &       &     & RV & $0.852 \pm 0.0884$ & $5.366 \pm 3.7532$ \\
\hline
MnM2      & LA ED & 3D  full. & LV & $0.968 \pm 0.0216$ & $2.977 \pm 2.2622$ \\
          &       &     & MYO & $0.878 \pm 0.0571$ & $2.151 \pm 1.6604$ \\
          &       &     & RV & $0.934 \pm 0.0302$ & $4.075 \pm 2.2530$ \\ \cmidrule{2-6}
          & LA ES & 3D  full. & LV & $0.948 \pm 0.0376$ & $3.246 \pm 3.1302$ \\
          &       &     & MYO & $0.891 \pm 0.0666$ & $2.809 \pm 4.2584$ \\
          &       &     & RV & $0.899 \pm 0.0656$ & $4.254 \pm 3.4218$ \\ \cmidrule{2-6}
          & SA ED & 2D & LV & $0.957 \pm 0.0272$ & $3.169 \pm 3.4790$ \\
          &       &     & MYO & $0.867 \pm 0.0546$ & $2.842 \pm 3.5519$ \\
          &       &     & RV & $0.934 \pm 0.0415$ & $4.083 \pm 3.8421$ \\ \cmidrule{3-6}
          &       & 3D  full. & LV & $0.955 \pm 0.0252$ & $3.571 \pm 3.5109$ \\
          &       &     & MYO & $0.862 \pm 0.0525$ & $2.561 \pm 2.4932$ \\
          &       &     & RV & $0.934 \pm 0.0403$ & $4.200 \pm 3.8857$ \\ \cmidrule{3-6}
          &       & Ens. & LV & $0.958 \pm 0.0253$ & $3.256 \pm 3.4665$ \\
          &       &     & MYO & $0.869 \pm 0.0518$ & $2.371 \pm 2.4556$ \\
          &       &     & RV & $0.937 \pm 0.0394$ & $4.021 \pm 4.1842$ \\ \cmidrule{2-6}
          & SA ES & 2D & LV & $0.958 \pm 0.0272$ & $3.170 \pm 3.4631$ \\
          &       &     & MYO & $0.867 \pm 0.0543$ & $2.571 \pm 2.8774$ \\
          &       &     & RV & $0.934 \pm 0.0425$ & $4.520 \pm 6.4278$ \\ \cmidrule{3-6}
          &       & 3D  full. & LV & $0.956 \pm 0.0252$ & $3.481 \pm 3.4478$ \\ 
          &       &     & MYO & $0.862 \pm 0.0521$ & $2.555 \pm 2.4800$ \\
          &       &     & RV & $0.934 \pm 0.0414$ & $4.302 \pm 4.0415$ \\ \cmidrule{3-6}
          &       & Ens. & LV & $0.958 \pm 0.0255$ & $3.264 \pm 3.4861$ \\ 
          &       &     & MYO & $0.868 \pm 0.0516$ & $2.341 \pm 2.4096$ \\
          &       &     & RV & $0.938 \pm 0.0380$ & $3.929 \pm 4.0974$ \\
\hline
\end{longtable}
\end{center}

\subsection{Comparison with other models}

\subsubsection{LAScarQS}

\begin{figure}[!h]
    \centering
    \includegraphics[width=0.8\linewidth]{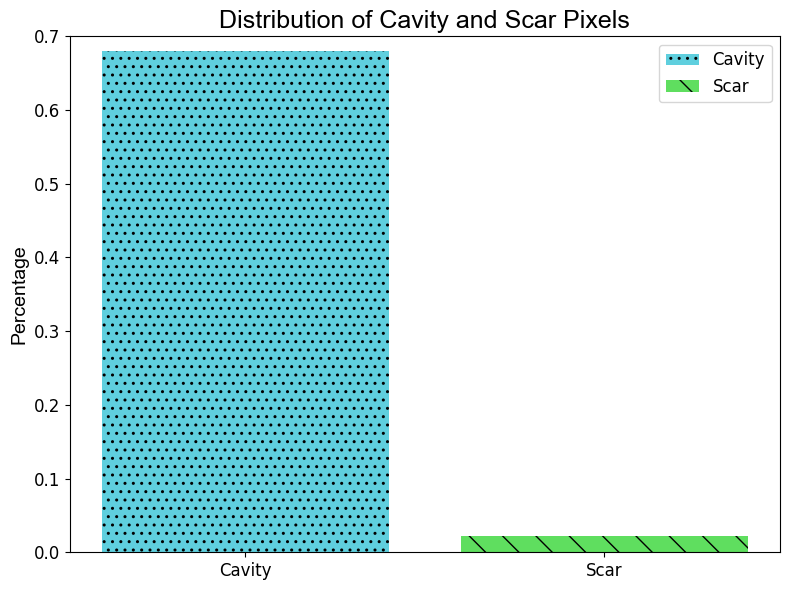}
    \caption{Total pixel distribution of 60 images of LAScarQS Task1. }
    \label{fig:sup_pix_dist}
\end{figure}

In Fig.~\ref{fig:sup_pix_dist} shows the pixel distribution of the cavity and scar for 60 MRIs from the LAScarQS Task 1. While most of the area belongs to the background, nearly 0.7\% pixels are cavity and less than 0.1\% belongs to the scar region. This showcases the class imbalance nature of the scar and cavity pixels.

In Table~\ref{tab:sup_lascar1_compare} we compare the results of the nnU-Net with other models which used the same dataset for Task1 using dice scores. Table~\ref{tab:sup_lascar1_compare} shows the performance of the Task 2 with nnUNet and other models with DSC, HD and MSD matrices.

% \begin{table}[!h]
% \caption{Performance comparison of dice scores in nnU-Net variations and other models in LAScarQS-Task1 for scar and cavity segmentation.}
% \label{tab:sup_lascar1_compare}
% \centering
% \setlength{\tabcolsep}{3pt} 
% \begin{tabular}{lcc} 
% \toprule
% \textbf{Paper}& \textbf{Scars} & \textbf{Cavity} \\ \midrule
% \cite{punithakumar2022automated} & 0.660& 0.907 \\
% \cite{jiang2022deep} & 0.641 & 0.902 \\ 
% \cite{arega2022using} & 0.634 & 0.898 \\ 
% \cite{mazher2022automatic} & 0.602 & 0.875 \\ 
% \cite{zhang2022automatically} & 0.598 & 0.880 \\
% \cite{lefebvre2022lassnet} & 0.553 & 0.938\\
% \midrule
%  nnU-Net (2D)
% & 0.439&0.926\\
%   nnU-Net (3D full res)
% & 0.443&0.939\\
%   nnU-Net (3D low res)
% & 0.411&0.937\\
%   nnU-Net (3D cascade)
% & 0.449&0.939\\
%   nnU-Net (Ensemble)& 0.439&0.939\\
% \bottomrule
% \end{tabular}
% \end{table}

\begin{table}[!h]
\caption{Performance comparison of dice scores in nnU-Net variations and other models in LAScarQS-Task1 for scar and cavity segmentation.}
\label{tab:sup_lascar1_compare}
\centering
\setlength{\tabcolsep}{3pt}
\begin{tabular}{lcc}
\toprule
\textbf{Method} & \textbf{Scars} & \textbf{Cavity} \\ \midrule
Punithakumar et al. (2022)\cite{punithakumar2022automated} & 0.660 & 0.907 \\
Jiang et al. (2022)\cite{jiang2022deep} & 0.641 & 0.902 \\
Arega et al. (2022)\cite{arega2022using} & 0.634 & 0.898 \\
Mazher et al. (2022)\cite{mazher2022automatic} & 0.602 & 0.875 \\
Zhang et al. (2022)\cite{zhang2022automatically} & 0.598 & 0.880 \\
Lefebvre et al. (2022)\cite{lefebvre2022lassnet} & 0.553 & 0.938 \\ \midrule
nnU-Net (2D) & 0.439 & 0.926 \\
nnU-Net (3D full resolution) & 0.443 & 0.939 \\
nnU-Net (3D low resolution) & 0.411 & 0.937 \\
nnU-Net (3D cascade) & 0.449 & 0.939 \\
nnU-Net (Ensemble) & 0.439 & 0.939 \\ \bottomrule
\end{tabular}
\end{table}

\begin{table}[!h]
\caption{Performance comparison of nnU-Net variations and other models in LAScarQS-Task2.  DSC- Dice Score, HD - Hausdorff Distance, MSD- Mean Surface Distance}
\label{tab:sup_lascar2_compare}
\centering
\setlength{\tabcolsep}{3pt}
\begin{tabular}{lccc}
\toprule
\textbf{Method} & \textbf{DSC} & \textbf{HD} & \textbf{MSD}\\ \midrule
Lefebvre et al. (2022)\cite{lefebvre2022lassnet} & 0.889 & 26.270 & 2.179 \\
Tu et al. (2022)\cite{tu2022self} & 0.890 & 17.124 & 1.706 \\
Liu et al. (2022)\cite{liu2022ugformer} & 0.866 & – & – \\
Zhang et al. (2022)\cite{zhang2022automatically} & 0.890 & 16.450 & 1.715 \\
Zhang et al. (2022)\cite{zhang2022two} & 0.878 & – & 0.710 \\
Khan et al. (2022)\cite{khan2022sequential} & 0.846 & 105.700 & 3.390 \\
Xie et al. (2022)\cite{xie2022hrnet} & 0.872 & 22.394 & – \\
Zhou et al. (2022)\cite{zhou2022edge} & 0.875 & 24.731 & 2.233 \\
Jiang et al. (2022)\cite{jiang2022deep} & 0.881 & 18.755 & 1.782 \\
Li et al. (2022)\cite{li2022cross} & 0.883 & 20.883 & 1.794 \\
Arega et al. (2022)\cite{arega2022using} & 0.890 & 16.907 & 1.720 \\
Punithakumar et al. (2022)\cite{punithakumar2022automated} & 0.893 & 15.860 & 1.613 \\
Mazher et al. (2022)\cite{mazher2022automatic} & 0.886 & 18.389 & 1.813 \\
Singh et al. (2023)\cite{singh2023attention} & 0.929 & 12.960 & 0.890 \\
Singh et al. (2023)\cite{singh2023madru} & 0.919 & 15.430 & - \\ \midrule
nnU-Net (2D) & 0.930 & 13.971 & 0.733 \\
nnU-Net (3D full resolution) & 0.937 & 12.971 & 0.672 \\
nnU-Net (3D low resolution) & 0.935 & 12.741 & 0.692 \\
nnU-Net (3D cascade) & 0.937 & 12.807 & 0.667 \\
nnU-Net (Ensemble) & 0.938 & 12.767 & 0.652 \\ \bottomrule
\end{tabular}
\end{table}

\subsubsection{LASC}
In Table~\ref{tab:sup_LASC_compare}, the performance of the nnU-Net and other models are compared using DSC values.

\begin{table}[!h]
\caption{Performance comparison of nnU-Net variations and other models in LASC.  DSC- Dice Score}
\label{tab:sup_LASC_compare}
\centering
\begin{tabular}{lc}
\toprule
\textbf{Publication} & \textbf{DSC} \\ \midrule
Xia et al. (2019)\cite{xia2019automatic} & 0.932 \\
Bian et al. (2018)\cite{bian2018pyramid} & 0.926 \\
Vesal et al. (2019)\cite{vesal2019dilated} & 0.925 \\
Yang et al. (2019)\cite{yang2019combating} & 0.925 \\
Li et al. (2019)\cite{li2019attention} & 0.923 \\
Chen et al. (2022)\cite{chen2022combining} & 0.920 \\
Chen et al. (2023)\cite{chen2023transformer} & 0.932 \\
Li et al. (2023)\cite{li2023comprehensive} & 0.919 \\
Liu et al. (2019)\cite{liu2019deep} & 0.903 \\
Borra et al. (2019)\cite{borra2019semantic} & 0.898 \\
Puybareau et al. (2018)\cite{puybareau2018left} & 0.923 \\
Uslu et al. (2021)\cite{uslu2021net} & 0.920 \\
Chen et al. (2021)\cite{chen2021jas} & 0.913 \\
Chen et al. (2022)\cite{chen2022multiresolution} & 0.923 \\
Qi et al. (2023)\cite{qi2023cardiac} & 0.921 \\
Zhao et al. (2023)\cite{zhao2023context} & 0.911 \\
Singh et al. (2023)\cite{singh2023attention} & 0.935 \\ 
Singh et al. (2023)\cite{singh2023madru} & 0.934 \\
Milletari et al. (2016)\cite{milletari2016v} & 0.919\\
Lourenço et al. (2021)\cite{lourencco2021left} & 0.910\\
Zhao et al. (2021)\cite{zhao2021not} & 0.918\\
Liu et al. (2022)\cite{liu2022uncertainty} & 0.920\\
Xu et al. (2024)\cite{xu2024dynamic} & 0.926\\
 \midrule
  nnU-Net (2D)               & 0.926 \\
 nnU-Net (3D full resolution)      & 0.933 \\
 nnU-Net (3D low resolution)       & 0.931 \\
 nnU-Net (3D cascade)       & 0.933 \\
 nnU-Net (Ensemble)         & 0.934 \\
\bottomrule
\end{tabular}

\end{table}

% \begin{table}[!h]
% \caption{Performance comparison of nnU-Net variations and other models in LASC.  DSC- Dice Score}
% \label{tab:sup_LASC_compare}
% \centering
% \begin{tabular}{lc}
% \toprule
% \textbf{Method} & \textbf{DSC} \\ \midrule
% \cite{xia2019automatic} & 0.932 \\
% \cite{bian2018pyramid} & 0.926 \\
% \cite{vesal2019dilated} & 0.925 \\
% \cite{yang2019combating} & 0.925 \\
% \cite{li2019attention} & 0.923 \\
% \cite{chen2022combining} & 0.920 \\
% \cite{chen2023transformer} & 0.932 \\
% \cite{li2023comprehensive} & 0.919 \\
% \cite{liu2019deep} & 0.903 \\
% \cite{borra2019semantic} & 0.898 \\
% \cite{puybareau2018left} & 0.923 \\
% \cite{uslu2021net} & 0.920 \\
% \cite{chen2021jas} & 0.913 \\
% \cite{chen2022multiresolution} & 0.923 \\
% \cite{qi2023cardiac} & 0.921 \\
% \cite{zhao2023context} & 0.911 \\
% \cite{singh2023attention} & 0.935 \\ 
% \cite{singh2023madru} & 0.934 \\
% \cite{milletari2016v}& 0.919\\
% \cite{lourencco2021left}& 0.910\\
% \cite{zhao2021not}& 0.918\\
% \cite{liu2022uncertainty}& 0.920\\
% \cite{xu2024dynamic}& 0.926\\
%  \midrule
%   nnU-Net (2D)               & 0.926 \\
%  nnU-Net (3D full res)      & 0.933 \\
%  nnU-Net (3D low res)       & 0.931 \\
%  nnU-Net (3D cascade)       & 0.933 \\
%  nnU-Net (Ensemble)         & 0.934 \\
% \bottomrule
% \end{tabular}

% \end{table}

\subsubsection{ACDC}

In Table~\ref{tab:sup_acdc_compare}, the performance of the nnU-Net and other models are compared using DSC and HD values for both End Diastole (ED) and End Systole (ES) phases for Left Ventricles (LV), Myocardium (MYO) and Right Ventricles (RV).

\begin{sidewaystable}[!h]
% \begin{table}[!h]
\caption{Performance comparison of nnU-Net variations and other models in ACDC. LV- Left Ventricle, MYO- Myocardium, RV- Right Ventricle, ED- End Diastole, ES - End Systole, DSC- Dice Score, HD - Hausdorff Distance.}
\label{tab:sup_acdc_compare}
\setlength{\tabcolsep}{3pt}
\centering
\begin{tabular}{l|cc|cc|cc|cc|cc|cc}
\hline
\multirow{3}{*}{Method} & \multicolumn{4}{c|}{LV} & \multicolumn{4}{c|}{MYO} & \multicolumn{4}{c}{RV} \\ \cmidrule{2-13}
& \multicolumn{2}{c|}{ED} & \multicolumn{2}{c|}{ES} & \multicolumn{2}{c|}{ED} & \multicolumn{2}{c|}{ES} & \multicolumn{2}{c|}{ED} & \multicolumn{2}{c}{ES} \\ \cmidrule{2-13}
& DSC & HD & DSC & HD & DSC & HD & DSC & HD & DSC & HD & DSC & HD \\ \midrule
Guo et al. (2021)\cite{guo2021cardiac} & 0.968 & 5.814 & 0.935 & 7.361 & 0.906 & 7.469 & 0.923 & 7.702 & 0.955 & 8.877 & 0.894 & 11.649 \\
Isensee et al. (2018)\cite{isensee2018automatic} & 0.967 & 5.476 & 0.928 & 6.921 & 0.904 & 7.014 & 0.923 & 7.328 & 0.951 & 8.205 & 0.904 & 11.665 \\
Simantiris et al. (2020)\cite{simantiris2020cardiac} & 0.967 & 6.366 & 0.928 & 7.573 & 0.891 & 8.264 & 0.904 & 9.575 & 0.936 & 13.289 & 0.889 & 14.367 \\
Berihu et al. (2021)\cite{berihu2021learning} & 0.968 & 6.422 & 0.916 & 9.305 & 0.894 & 8.998 & 0.906 & 9.922 & 0.939 & 11.326 & 0.893 & 13.306 \\
Ammar et al. (2021)\cite{ammar2021automatic} & 0.968 & 7.993 & 0.911 & 10.528 & 0.891 & 10.575 & 0.901 & 13.891 & 0.929 & 14.189 & 0.886 & 16.042 \\
Zotti et al. (2018)\cite{zotti2018convolutional} & 0.964 & 6.180 & 0.912 & 8.386 & 0.886 & 9.586 & 0.902 & 9.291 & 0.934 & 11.052 & 0.885 & 12.650 \\
Khened et al. (2018)\cite{khened2018densely} & 0.964 & 8.129 & 0.917 & 8.968 & 0.889 & 9.841 & 0.898 & 12.582 & 0.935 & 13.994 & 0.879 & 13.930 \\
Baumgartner et al. (2018)\cite{baumgartner2018exploration} & 0.963 & 6.526 & 0.911 & 9.170 & 0.892 & 8.703 & 0.901 & 10.637 & 0.932 & 12.670 & 0.883 & 14.691 \\
Painchaud et al. (2020)\cite{painchaud2020cardiac} & 0.961 & 6.152 & 0.911 & 8.278 & 0.881 & 8.651 & 0.897 & 9.598 & 0.933 & 13.718 & 0.884 & 13.323 \\
Wolterink et al. (2018)\cite{wolterink2018automatic} & 0.961 & 7.515 & 0.918 & 6.603 & 0.875 & 11.121 & 0.894 & 10.687 & 0.928 & 11.879 & 0.872 & 13.399 \\
Calisto et al. (2020)\cite{calisto2020adaen} & 0.958 & 5.592 & 0.903 & 8.644 & 0.873 & 8.197 & 0.895 & 8.318 & 0.936 & 10.183 & 0.884 & 12.234 \\
Zotti et al. (2018)\cite{zotti2018gridnet} & 0.957 & 6.641 & 0.905 & 8.706 & 0.884 & 8.708 & 0.896 & 9.264 & 0.941 & 10.318 & 0.882 & 14.053 \\
Singh et al. (2023)\cite{singh2023w} & 0.967 & 5.526 & 0.935 & 6.913 & 0.902 & 8.094 & 0.921 & 7.772 & 0.949 & 9.187 & 0.900 & 11.556 \\
Singh et al. (2023)\cite{singh2023attention} & 0.967 & 5.652 & 0.938 & 6.878 & 0.905 & 7.389 & 0.923 & 7.373 & 0.950 & 8.513 & 0.895 & 12.167 \\
Singh et al. (2023)\cite{singh2023madru} & 0.968 & 5.859 & 0.937 & 6.529 & 0.904 & 7.723 & 0.922 & 7.221 & 0.952 & 8.788 & 0.890 & 11.926 \\ \midrule
% nnU (2D) & 0.942 & 10.438 & 0.885 & 12.678 & 0.897 & 10.050 & 0.913 & 8.231 & 0.965 & 6.739 & 0.927 & 6.795 \\
% nnU (3D full resolution) & 0.934 & 11.494 & 0.882 & 12.743 & 0.889 & 8.057 & 0.906 & 8.785 & 0.959 & 8.486 & 0.901 & 9.028 \\
% nnU (Ensemble) & 0.944 & 10.716 & 0.892 & 12.200 & 0.898 & 9.884 & 0.915 & 8.460 & 0.963 & 9.584 & 0.922 & 8.321 \\ \bottomrule

nnU (2D) &  0.965 & 6.739 & 0.927 & 6.795 & 0.897 & 10.050 & 0.913 & 8.231  & 0.942 & 10.438 & 0.885 & 12.678 \\
nnU (3D full resolution) & 0.959 & 8.486 & 0.901 & 9.028 & 0.889 & 8.057 & 0.906 & 8.785 & 0.934 & 11.494 & 0.882 & 12.743  \\
nnU (Ensemble) & 0.963 & 9.584 & 0.922 & 8.321  & 0.898 & 9.884 & 0.915 & 8.460 &    0.944 & 10.716 & 0.892 & 12.200 \\ \bottomrule
\end{tabular}
\end{sidewaystable}

\subsubsection{MnM}

In Table~\ref{tab:sup_mnm1_compare}, the performance of the nnU-Net and other models are compared using DSC and HD values for both ED and ES phases for LV, MYO and RV.

\begin{sidewaystable}[!h]
    \caption{Performance comparison of nnU-Net variations and other models in MnM. LV- Left Ventricle, MYO- Myocardium, RV- Right Ventricle, ED- End Dystole, ES - End Systole, DSC- Dice Score, HD - Hausdorff Distance.}
    \label{tab:sup_mnm1_compare}
    \centering
    \begin{tabular}{l|cc|cc|cc|cc|cc|cc}
    \toprule
Method   &\multicolumn{4}{|c|}{LV}&  \multicolumn{4}{c}{MYO}&  \multicolumn{4}{c}{RV}\\ \cmidrule{2-13}
          &\multicolumn{2}{|c|}{ED}&  \multicolumn{2}{c|}{ES}&  \multicolumn{2}{c|}{ED}&  \multicolumn{2}{c|}{ES}&  \multicolumn{2}{c|}{ED}& \multicolumn{2}{c}{ES}\\ \cmidrule{2-13}
          &DSC&  HD&  DSC&  HD&  DSC&  HD&  DSC&  HD&  DSC& HD & DSC&HD \\ \midrule
Full et al. (2021)\cite{full2021studying} & 0.939 & 9.1 & 0.886 & 9.1 & 0.839 & 12.8 & 0.867 & 10.6 & 0.910 & 11.8 & 0.860 & 12.7 \\
Parreno et al. (2021)\cite{parreno2021deidentifying} & 0.939 & 11.3 & 0.884 & 11.4 & 0.826 & 15.2 & 0.856 & 14.0 & 0.886 & 15.4 & 0.829 & 16.7 \\
Zhang et al. (2021)\cite{zhang2021semi} & 0.938 & 9.3 & 0.880 & 9.5 & 0.830 & 12.9 & 0.861 & 10.8 & 0.909 & 12.3 & 0.850 & 13.0 \\
Ma et al. (2021)\cite{ma2021histogram} & 0.935 & 9.5 & 0.875 & 10.5 & 0.825 & 13.3 & 0.856 & 11.6 & 0.906 & 12.3 & 0.844 & 13.0 \\
Saber et al. (2021)\cite{saber2021multi} & 0.933 & 13.4 & 0.867 & 14.0 & 0.812 & 17.1 & 0.839 & 18.2 & 0.876 & 15.7 & 0.815 & 18.1 \\
Kong et al. (2021)\cite{kong2021generalizable} & 0.931 & 10.0 & 0.877 & 9.8 & 0.816 & 13.7 & 0.850 & 11.3 & 0.893 & 14.3 & 0.827 & 15.2 \\
Singh et al. (2023)\cite{singh2023w} & 0.928 & 7.15 & 0.890 & 7.6 & 0.834 & 10.2 & 0.868 & 9.6 & 0.902 & 10.6 & 0.852 & 11.7 \\
Corral et al. (2021)\cite{corral20212} & 0.927 & 11.2 & 0.877 & 9.7 & 0.815 & 14.0 & 0.852 & 11.1 & 0.892 & 13.6 & 0.834 & 15.0 \\
Li et al. (2021)\cite{li2021generalisable} & 0.922 & 15.5 & 0.857 & 17.5 & 0.809 & 18.0 & 0.836 & 17.2 & 0.867 & 16.6 & 0.802 & 19.1 \\
Khader et al. (2021)\cite{khader2021adaptive} & 0.914 & 12.1 & 0.853 & 12.0 & 0.768 & 17.2 & 0.814 & 15.2 & 0.850 & 17.5 & 0.794 & 17.0 \\
Carscadden et al. (2021)\cite{carscadden2021deep} & 0.913 & 14.5 & 0.851 & 13.0 & 0.776 & 17.8 & 0.809 & 14.5 & 0.791 & 30.7 & 0.732 & 32.9 \\
Scannell et al. (2021)\cite{scannell2021domain} & 0.905 & 13.6 & 0.848 & 15.5 & 0.772 & 17.2 & 0.820 & 17.5 & 0.876 & 16.2 & 0.809 & 19.6 \\
Huang et al. (2021)\cite{huang2021style} & 0.896 & 15.7 & 0.772 & 23.0 & 0.761 & 17.9 & 0.721 & 20.2 & 0.820 & 21.0 & 0.698 & 29.5 \\
Liu et al. (2021)\cite{liu2021disentangled} & 0.889 & 16.0 & 0.835 & 14.2 & 0.785 & 22.1 & 0.808 & 18.9 & 0.814 & 22.1 & 0.758 & 22.0 \\
Li et al. (2021)\cite{li2021random} & 0.797 & 21.9 & 0.716 & 25.8 & 0.668 & 31.6 & 0.673 & 33.0 & 0.552 & 49.1 & 0.517 & 52.0 \\
Singh et al. (2023)\cite{singh2023attention} & 0.940 & 7.5 & 0.890 & 7.7 & 0.839 & 10.3 & 0.870 & 9.9 & 0.909 & 10.2 & 0.856 & 11.4 \\ \midrule

nnU (2D)        & 0.936 & 7.5 & 0.888 & 12.7 & 0.824 & 10.7 & 0.800 & 15.4 & 0.909 & 11.6 & 0.893  & 14.6\\
nnU (3D full resolution)        & 0.933 & 8.2 & 0.909 & 8.6  & 0.819 & 10.8 & 0.841 & 11.1 & 0.908 &11.5 & 0.871 & 13.1 \\
nnU (Ensemble)  & 0.937 & 7.4 & 0.888 & 8.5  & 0.826 & 9.9 & 0.864 & 9.9 & 0.913  & 10.8 & 0.852 & 12.7 \\
\bottomrule
\end{tabular}

\end{sidewaystable}

\subsubsection{MnM2}

In Table~\ref{tab:MnM2_compare}, the performance of the nnU-Net and other models are compared using DSC and HD values for both ED and ES phases for RV only for both short and long axis.

\begin{sidewaystable}[!h]
    \caption{Performance comparison of nnU-Net variations and other models in MnM2 for Right Ventricle only. ShA- Short Axis, LoA - Long Axis, ED - End Diastole, ES- End Systole, DSC - Dice Score, HD - Hausdorff Distance.}
    \label{tab:MnM2_compare}
    \centering
    \begin{tabular}{l|cc|cc|cc|cc}
        \toprule
        & \multicolumn{4}{c}{ShA}& \multicolumn{4}{c}{LoA}\\ \cmidrule{2-9}
        Method & \multicolumn{2}{c|}{ED} & \multicolumn{2}{c|}{ES} & \multicolumn{2}{c|}{ED} & \multicolumn{2}{c}{ES} \\ \cmidrule{2-9}

        & DSC & HD & DSC & HD & DSC & HD & DSC & HD \\

        \midrule
        Fulton et al. (2021)\cite{fulton2021deformable} & 0.934 & 9.610& 0.910 & 10.032 & 0.935& 6.227 & 0.904 & 5.935\\
        Arega et al. (2021)\cite{arega2021using} & 0.932 & 10.078 & 0.910 & 9.782& 0.935 & 6.028& 0.905& 6.188 \\
        Punithakumar et al. (2021)\cite{punithakumar2021automated} & 0.940& 10.122 & 0.914& 9.987 & 0.931 & 6.337 & 0.904 & 5.976 \\
        Li et al. (2021)\cite{li2021right} & 0.933 & 10.563 & 0.907 & 10.050 & 0.930 & 6.246 & 0.902 & 6.097 \\
        Sun et al. (2022)\cite{sun2022right} & 0.937 & 10.879 & 0.913 & 9.874 & 0.935 & 6.056 & 0.904 & 6.031 \\
        Al et al. (2021)\cite{al2021late} & 0.927 & 9.941 & 0.897 & 10.307 & 0.907 & 8.444 & 0.883 & 7.265 \\
        Liu et al. (2021)\cite{liu2021refined} & 0.932 & 10.517 & 0.903 & 10.101 & 0.934 & 7.721 & 0.896 & 6.019 \\
        Jabbar et al. (2021)\cite{jabbar2021multi} & 0.923 & 11.258 & 0.897 & 11.062 & 0.910 & 7.757 & 0.882 & 6.933 \\
        Queiros et al. (2021)\cite{queiros2021right} & 0.924 & 11.327 & 0.898 & 11.447 & 0.922 & 7.173 & 0.900 & 6.391 \\
        Galati et al. (2021)\cite{galati2021using} & 0.916 & 11.681 & 0.890 & 11.747 & 0.924 & 7.840 & 0.894 & 6.978 \\
        Mazher et al. (2021)\cite{mazher2021multi} & 0.909 & 15.275 & 0.880 & 14.606 & 0.888 & 8.333 & 0.854 & 8.347 \\
        Gao et al. (2022)\cite{gao2022consistency} & 0.844 & 15.495 & 0.821 & 16.750 & 0.887 & 9.733 & 0.851 & 9.659 \\
        Beetz et al. (2021)\cite{beetz2021multi} & 0.873 & 16.682 & 0.820 & 17.913 & 0.896 & 8.570 & 0.864 & 7.591 \\
        Tautz et al. (2021)\cite{tautz20213d} & 0.883 & 17.024 & 0.838 & 18.003 & 0.849 & 13.303 & 0.809 & 13.716 \\
        Galazis et al. (2021)\cite{galazis2021tempera} & 0.852 & 19.430 & 0.821 & 19.117 & 0.814 & 18.629 & 0.781 & 17.198 \\ \midrule

         nnU (2D) & 0.934 & 10.50 & 0.934 & 11.228 & - & - & - & - \\
         nnU (3D full resolution) & 0.934 & 10.393 & 0.934 & 10.301 & 0.934 & 6.055 & 0.900 & 6.108 \\
         nnU (Ensemble) & 0.937 & 11.079 & 0.938 & 11.119 & - & - & - & - \\
        \bottomrule
    \end{tabular}
\end{sidewaystable}

% \newpage
% \bibliography{sn-bibliography_sup}
% \bibliography{sn-bibliography}

\end{document}